\documentclass[twocolumn,aps,prb,showpacs]{revtex4}
\usepackage{makeidx}
\usepackage{amssymb}
\usepackage{amssymb,amsmath}
\usepackage{graphicx}
\usepackage{graphics}
\usepackage{dsfont}

\newcommand{\Tr}{\mathrm{Tr}}

\newcommand{\PathToFigures}{}

\begin{document}
\title{ Two dimensional Dirac fermions in the presence of long-range
correlated disorder  }
\author{Andrei A. Fedorenko, David Carpentier and Edmond Orignac  }
\affiliation{ CNRS UMR5672 -- Laboratoire de Physique de l'Ecole Normale
Sup{\'e}rieure de Lyon, 46, All{\'e}e d'Italie, 69007 Lyon, France }

\date{December 13, 2011}

\pacs{73.63.-b,73.22.Pr,73.23.-b}

\begin{abstract}
We consider 2D Dirac fermions in the presence of three types of disorder:
random scalar potential, random gauge potential and random mass with long-range
correlations decaying as a power law. Using various methods such as the
self-consistent Born approximation (SCBA), renormalization group (RG), the
matrix Green function formalism and bosonisation we calculate the density
of states and study the full counting statistics of fermionic transport at
lower energy. The SCBA and RG show that the random correlated scalar potentials
generate an algebraically small energy scale below which the density
of states saturates to a constant value.
For correlated random gauge potential, RG and bosonisation calculations provide
consistent behavior
of the density of states which diverges at zero energy in an integrable way.
In the case of correlated random mass disorder the RG flow has a nontrivial
infrared stable  fixed point leading to a universal power-law behavior of the
density of states and also to universal transport properties. In contrast to
uncorrelated  case the correlated scalar potential and random mass disorders
give rise to deviation from the pseudodiffusive transport already to
lowest order in disorder strength.
\end{abstract}

\maketitle

\section{Introduction}

In the recent years significant attention has been attracted by materials
exhibiting two dimensional fermionic excitations with linear dispersion
relation near the Fermi level. These excitations share many properties with
massless relativistic particles but with a velocity reduced with respect to
the speed of light. The seminal example of such material is
graphene,\cite{novoselov05} the low energy properties of which are described
by two dimensional (2D) gas of Dirac fermions.\cite{castroneto09,dassarma11}
More recently, 2D Dirac fermions have also emerged as the effective low energy
degree of freedom in the surface states of 3D topological
insulators,\cite{fu07,moore07,roy09} such as the materials in the
Bi$_{2}$Se$_{3}$ family\cite{hsieh08}  and strained HgTe.\cite{hancock11}
Dirac excitations have been also found in unconventional superconductor
with $d$-wave symmetry\cite{orenstein00,asboth11,wang11,durst00},
in the quasi-2D organic conductor $\alpha$-(BEDT-TTF)$_{2}$ I$_{3}$ under
pressure,\cite{katayama06,kobayashi07,fukuyama07,goerbig08,nishine10,kobayashi11}
and even in photonic crystals.\cite{huang11}

The peculiar features of Dirac fermions lead to unfamiliar transport properties
in all these materials. Of particular interest
are the vanishing of the density of states at the Dirac (or neutrality) point,
and the unusual scattering properties
of the Dirac particles, a striking example of which being provided by
the so-called
Klein tunneling phenomenon:
an excitation incident normal to a potential barrier
crosses this barrier completely even for energies smaller than the barrier
height.\cite{castroneto09} A consequence of that is the total absence
of backscattering for these Dirac fermions. The peculiarities of the
spectrum near the Dirac point also affect transport strongly.
In undoped graphene, due to the evanescent nature of
the states at the Dirac point, transport in a clean sample is similar to
that in a diffusive wire.\cite{Tworzydlo06}  Moreover, all the different
cumulants of current
fluctuations in a graphene sample behave as in a
diffusive wire rather than in a ideal metallic system. In particular,
pseudodiffusive conductance scales with the length of the system $L$
as $1/L$ in contrast with the $L^0$ scaling  of ideal conducting systems.
The associated Fano factor defined as the ratio between the shot noise power
and the current has the same universal value $F=1/3$ as for diffusive metallic
wires instead of $F=0$ for ideal conductors.
Remarkably, the conductivity minimum of graphene at the neutrality point
due to evanescent modes is of order $e^2/h$, i.e. finite despite the vanishing
of the density of states. Though the conductivity minimum remains almost
constant in very broad temperature range its sample-dependance indicates the
importance of disorder for the transport properties of graphene.\cite{mucciolo10}

Due to these specificities, numerous studies have
focus on the effect of a random scattering
potential  on transport of Dirac
states.\cite{ostrovsky06,khveshchenko07,schuessler09,schuessler10}
Indeed, various kinds of disorder are naturally present in real materials, and
affect in a dominant way the electronic transport properties.
They can be of different origin: lattice defects, impurities, ripples
in graphene
sheet that distort locally the lattice, adatoms deposited on the surface of a
graphene sample,\cite{shevtsov11} atomic steps on the surface of topological
insulators, etc. Theoretical research on disordered
2D Dirac fermions was also  motivated initially by its relevance to
Quantum Hall transitions.\cite{ludwig94} Building on
this pioneering work, recent studies have focused on the effect of the
different types of disorder on the transport properties of Dirac fermions as in
the limit of low energy, i.e. around the Dirac point, as well as away from half
filling.
In the present paper, we do not consider the highly doped weak localization
regime\cite{maccann06,kharitonov08} corresponding to $k_{F}l_{0}\gg 1$, where $l_{0}$
is the mean free path, and concentrate mostly on the transport near
the Dirac point where $k_F \to 0$.
It has been shown that the conductivity
at half filling depends not only on the type
of disorder but also on the infrared cutoff, so that it potentially can depend
on the geometry of  the physical setup.\cite{ostrovsky06} The role of
 infrared cutoff can be played by either the mean free path,
the Fermi length or the size of the system. These different cases allow to
identify several transport
regimes.\cite{schuessler09} Fixing geometry, for example to
wide-and-short rectangle
with many propagating transverse modes one can compute the conductance and
the Fano factor for each regime.\cite{schuessler10}
Most of the previous studies addressed the case where the
scattering potential is uncorrelated in space. This is the case for
instance if it is  originated from localized point-like scatterers distributed
independently from each other.
However, several physically relevant types of disorder sources exhibit
long-range correlations.

For example, a graphene sheet is known to develop
static shape fluctuations due to the
unavoidable thermodynamic instability of 2D crystals with respect to
both crumpling and bending. These ripples survive at low temperatures and can be
viewed as a static random gauge potential playing the role of a
quenched disorder on the  electronic time  scale (see below).
The theory of 2D elastic membranes predicts the strength of the local height
fluctuations which give rise to long-range algebraic correlation of this random
gauge potential.\cite{katnelson07,fasolino07,khveshchenko07,abedpour07,guinea08}
A second example, in the case of the surface states of topological
insulators, is surface roughness\cite{alpichshev11} which is one of
the dominant form of disorder in these materials.
A typical roughness created by atomic steps can induce a scattering potential
with algebraically decaying  correlations.\cite{nozieres87,fedorenko06}
As a last example, the adsorption of magnetic adatoms on the surface of
 topological insulators has been proposed as a way to control the
electronic properties of the surface states.\cite{abanin11}
If the characteristic
spin flip time of magnetic adatoms exceeds the mean free time of the electrons
in the surface states these adatoms can be also viewed as a source of quenched
disorder of both random gauge and random mass types.
In the vicinity of the paramagnetic-ferromagnetic transition, induced by the
RKKY-type surface interactions, critical magnetization fluctuations
will give rise to a quenched disorder on electronic time scales with
power law correlations in space.

Motivated by these physical examples, in this paper we consider  the general
properties of 2D Dirac fermions in the presence of various
weak random potentials possessing
algebraic spatial correlations. Using several analytical techniques,
we consider
perturbatively these disorder potentials. We  focus on the effect
of these long-range correlations on the density of states and
also transport properties  discussing the
cases relevant for the three above mentioned  examples.
In particular, we will compute the unknown to our knowledge
density of states for  correlated random potential and random mass.
We will show that the previous estimation\cite{khveshchenko08} of
the density of states for correlated random gauge potential is wrong. We
will find the correct density of states using two different methods:
renormalization group and bosonisation.
We will develop a framework to study transport properties
in the presence of correlated disorder using the matrix Green
function formalism introduced by Nazarov.\cite{nazarov94,ryu07}
Our approach goes beyond the previous work of
Khveshchenko\cite{khveshchenko07,khveshchenko08} who also considered
 LR correlated potentials but focused on
the multi fractal spectrum of wave functions at the Dirac points and
the conductance within the self consistant Born approximation (SCBA).

The paper is organized as follows. Section \ref{sec:model}
introduces the model. In Sec.~\ref{sec:SCBA} we consider the SCBA approximation.
In Sec.~\ref{sec:full} using the matrix Green function formalism we study
the full counting statistics for a wide-and-short rectangle sample
at the neutrality point.  In Sec.~\ref{sec:RG} we derive
RG equations to one-loop order and discuss the  properties of the
systems with different type of disorder. In Sec.~\ref{sec:bosonization}
we use bosonisation technique for systems with LR correlated random
gauge potential.
In Sec.~\ref{sec:summary} we summarize the obtained results.

\section{Model} \label{sec:model}

\subsection{Single flavor Dirac model}
Whenever they appear in a 2D or quasi-2D material, like graphene,
lattice Dirac
fermions are constrained by the Nielsen-Ninomiya theorem\cite{nielsen81} to
appear by pair of species. Practically, this implies the existence of
an even  number of Dirac cones in the first Brillouin zone when considering the
low energy  dispersion relation. Indeed, in graphene two Dirac cones exist
at the inequivalent points $K$ and $K'=-K$ at the zone boundary.  However,
any potential varying on scales much larger that the atomic
scale $\pi /K$ will leave the two Dirac cones uncoupled.
In this case, the effect of the
potential can be described by considering its effect on a single Dirac cone,
treating the presence of the other Dirac point as an effective degeneracy.
We thus lead to consider a single species of non-interacting massless 2D
Dirac fermions in the presence of a random potential,
described by the Hamiltonian
\begin{equation}\label{eq:Dirac1}
H=H_0+V(x,y),
\end{equation}
where $H_0$ is the kinetic Hamiltonian of free Dirac fermions with
the Fermi velocity $v_0$,
\begin{equation}\label{eq:Dirac2}
H_0 = -i v_0 (\sigma_x\partial_x+\sigma_y\partial_y),
\end{equation}
and $V(x,y)$ is a random disorder potential.
Here and in the following $\sigma_{0}=\mathds{1}$,
$\sigma_{\mu}$, $\mu=x,y,z$ are the respective Pauli matrices,
and  we set $\hbar=1$ for convenience.
This type of potential, without any Fourier component which couple
the different Dirac species, is often denoted as a long-range potential.
This notation, which refers to correlations at the scale of the
lattice space, should not be confused with the long-range (LR) correlation in space
on which we focus in this paper. The  latter  characterizes the long
distance $q\simeq 0$ behavior of the random potential correlations.
Note that in the case of graphene, as well as in the
quasi-2D $\alpha$-(BEDT-TTF)$_{2}$ I$_{3}$, the Pauli matrices entering the
relativistic kinetic Hamiltonian refer to a pseudo-spin describing the
relative weight of the electronic wave function on two sub-lattices.
Thus, the coupling of the random potential to these two sub lattices will be
reflected in the parametrization introduced below for this potential in
terms of these Pauli matrices.

The case of surface states of topological insulators is
different.\cite{hasan10,qi11} In
these materials, a strong spin orbit interaction opens a gap for
bulk states. The
non trivial topological order characterizing the filled bands of this
insulator implies the existence of Dirac fermions surface states.
Since they are not constrained by the
Nielsen-Ninomiya theorem, they occur around an odd number of Dirac
points in the first Brillouin zone.
In the simplest topological insulators, a single
Dirac cone exists at the surface of these insulators, the properties
of which control the surface transport properties of the material.
In this case, the
relativistic kinetic Hamiltonian of the surface states contains a real
magnetic term reflecting the bulk spin orbit interaction.
This term should be an
odd function of the electron spin, which we take for simplicity as the
in plane component of the spin, obtaining effectively
Eq.~(\ref{eq:Dirac1}). In this case, the parametrization of disorder
depends on the coupling of the corresponding potential to the spin
of the electrons.  In topological insulators,
the bulk topological order at the
origin of this odd number of  Dirac species, also prevents
time-reversal invariant disorder from
localizing these states.
This robustness property is in fact the result  of an
odd number of Dirac species as opposed to an even number, and will
play no role in the decoupled cones treatment of disorder that we will
perform.

As discussed above, we parametrize the disorder potential using the
following decomposition
\begin{equation} \label{eq:Disorder}
V(x,y)=\sum\limits_{\mu=0,x,y,z}\sigma_{\mu}V_{\mu}(x,y).
\end{equation}
In the usual terminology of disordered graphene the term with $\mu=0$
is called random potential disorder, the terms with $\mu=x,y$  random
gauge disorder and the term with
$\mu=z$  random mass disorder. We will keep this terminology, even though
for systems with real spin (instead of pseudo-spin) like topological
insulators they may have a different physical interpretation.
Fox instance, in topological insulators the terms with $\mu=x,y,z$ in
Eq.~(\ref{eq:Disorder}) correspond to real
random magnetic impurities on the conducting surface.

In what follows the disorder potentials $V_{\mu}(r)$ [$\mathbf{r}=\{x,y\}$]
are taken to be random and Gaussian with $\langle V_{\mu}(\mathbf{r}) \rangle=0$
and correlators
\begin{eqnarray}
\langle V_{\mu}(\mathbf{r})V_{\nu}(\mathbf{r}') \rangle =
2\pi  v_0^2\delta_{\mu \nu} g_{\mu}(\mathbf{r}-\mathbf{r}').
\end{eqnarray}
For the sake of convenience we fix the form of the correlator in Fourier space
\begin{eqnarray}
\langle V_{\mu}(\mathbf{k})V_{\nu}(\mathbf{k}') \rangle =(2\pi)^3 v_0^2
\delta(\mathbf{k}+\mathbf{k}') \delta_{\mu \nu}
(\alpha_{\mu}+\beta_{\mu}|\mathbf{k}|^{a-2}).
\end{eqnarray}
This form of the correlator corresponds in real space to
\begin{equation} \label{eq:g-mu}
 g_{\mu}(\mathbf{r})=\alpha_{\mu}\delta(\mathbf{r}) +
 \beta_{\mu} \mathcal{A}_a |\mathbf{r}|^{-a}, \ \ \
 \mathcal{A}_a=\frac{2^a\, \Gamma(a/2)}{4 \pi \Gamma(1-a/2)}
\end{equation}
The exponent $a$ is determined by the nature of disorder correlations
or by internal or fractal dimension of the extended defects.
In the presence of extended defects of internal dimension $\varepsilon_d$
 randomly orientated the corresponding exponent $a=2-\varepsilon_d$.
For instance, the presence linear dislocations or atomic steps with random
orientations on the surface of topological insulator ($\varepsilon_d=1$)
leads to LR correlated disorder with $a=1$.\cite{fedorenko06}
In the case of ripples of a graphene sheet, an evaluation of the correlation
of the generated gauge potential provides
an exponent\cite{katnelson07,fasolino07,khveshchenko07,abedpour07,guinea08}
 $a\approx1.6$.
In the case of magnetic adatoms deposited at the surface of topological
insulators,\cite{abanin11} the induced magnetic disorder correlations will decay
as a power law with $a=\eta$  where $\eta=1/4$ is the critical exponent
describing the magnetization correlation function in 2D Ising system.
Hence, these examples provide strong motivation to consider the effect of
these algebraic correlations  beyond the standard case with short-range correlation,
like $\delta$-corelation formally corresponding to $a>2$ in two dimensional systems.

\subsection{The effective replicated action}

For non-interacting fermions, we can write the partition function as:\cite{negele_book}
\begin{eqnarray}
Z = \int D\bar{\psi}(\mathbf{r},\tau) D\psi(\mathbf{r},\tau) e^{-S[\psi,\bar{\psi}]}
\end{eqnarray}
with the action given by
\begin{eqnarray}
S[\psi,\bar{\psi}] =   \int_0^\beta d\tau \int d^2 \mathbf{r}\, \bar{\psi}
(\mathbf{r},\tau)
 \left (\partial_\tau -\mu + H \right) \psi  (\mathbf{r},\tau) ,
\end{eqnarray}
where $\psi$ and $\bar{\psi}$ are anticommuting Grassman variables
satisfying the antisymmetric boundary condition $\psi(\mathbf{r},\tau+\beta)=-\psi(\mathbf{r},\tau)$
and  $\bar{\psi}(\mathbf{r},\tau+\beta)=-\bar{\psi}(\mathbf{r},\tau)$, and $\mu$ is the chemical
potential. For a time independent Hamiltonian, we can introduce the Fourier
series decomposition:
\begin{eqnarray}
  \label{eq:fourier}
   \psi  (\mathbf{r},\tau) = \frac 1 \beta \sum_{n} \psi(\mathbf{r},i\nu_n)
   e^{-i\nu_n \tau},
\end{eqnarray}
where $\nu_n=(2n+1)\pi/\beta$. This allows one to factorize the
partition function in such a way that all terms with different values of
$\nu_n$ are decoupled:
\begin{eqnarray}
  \label{eq:partition-prod}
  Z &=& \prod_{n} Z(i\nu_n), \ \   \\
  Z(i\nu_n) &=&  \int {\mathcal D} \bar{\psi} {\mathcal D}  \psi \  e^{-S(i\nu_n)}
 \nonumber \\
   &=&\mathrm{Det}(i\nu_n + \mu -H), \ \ \   \\
 S(i\nu_n)  &=&
\frac 1 \beta  \int d^2 \mathbf{r}\, \bar{\psi} (\mathbf{r},-i\nu_n) \left (H-i \nu_n -\mu\right)
    \psi  (\mathbf{r},i\nu_n). \ \ \ \ \
\end{eqnarray}
Hence, for each
value of $\nu_n$ we have a two-dimensional time independent field theory
in which the Matsubara frequency $\nu_n$ plays the role of a mass term.
Taking the zero temperature limit $\beta\to\infty$ one converts the sums
over $n$ into integrals over $\nu$ so that we end up with
\begin{eqnarray}
  \label{eq:partition-2}
  Z =  \int {\mathcal D} \bar{\psi} {\mathcal D} \psi \ \exp\left[
  -\int \frac{d \nu}{2\pi} S(i\nu)\right].
\end{eqnarray}
We are interested mostly in the properties of the undoped system
with the Fermi energy near the Dirac cone so that we put $\mu=0$
in what follows.
Using the replica trick\cite{edwards_replica} we derive the
replicated action for the fermions
at a given energy $\varepsilon=i\nu$. To that end we introduce $n$ replicas
of the original system and average their joint partition function over disorder
we obtain the effective action
\begin{eqnarray} \label{eq:action}
S(\varepsilon) & =& \sum\limits_{\alpha=1}^n
\int d^2 r \bar{\psi}_{\alpha}( \varepsilon +i v_0 \sigma_x\partial_x+i v_0
\sigma_y\partial_y)\psi_{\alpha}
\nonumber \\
&& \!\!\!\!\!\!\!\!
 + \pi v_0^2 \sum\limits_{\alpha,\beta=1}^n \sum\limits_{\mu=0}^3
\int d^2 \mathbf{r} \int d^2 \mathbf{r}' g_{\mu}(\mathbf{r}-\mathbf{r}') \nonumber \\
&&\!\!\!\!\!\!\!\!
 \times  [\bar{\psi}_{\alpha}(\mathbf{r})\sigma_{\mu} \psi_{\alpha}(\mathbf{r})]
[\bar{\psi}_{\beta}(\mathbf{r}')\sigma_{\mu} \psi_{\beta}(\mathbf{r}')].
\end{eqnarray}
The properties of the original system with quenched disorder are then obtain by taking
the limit $n\to 0$.

\section{Self-consistent Born approximation}
\label{sec:SCBA}

\begin{figure}
\includegraphics[width=8cm]{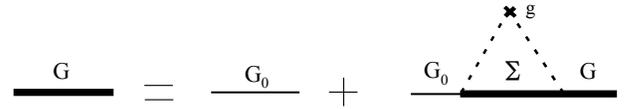}
\caption{The averaged over disorder Green function in the SCBA.}
  \label{fig:SCBA}
\end{figure}

Let us first consider the  self-consistent Born approximation (SCBA) which
is applicable only in the limit of  weak scattering. The SCBA has been widely
used to study the effect of uncorrelated disorder\cite{lee93,ostrovsky06,fukuzawa09}
as well as the effect of Coulomb impurities\cite{khveshchenko07} on the Dirac
fermions.
The retarded and advanced Green functions can be expressed in terms of
the self-energy via the Dyson equation
\begin{eqnarray}
G(\varepsilon) = G_0(\varepsilon) + G_0(\varepsilon) \Sigma(\varepsilon)
G(\varepsilon), \label{eq:SCBA}
\end{eqnarray}
where
$G_0(\varepsilon,\mathbf{k})=(\varepsilon-v_0 \boldsymbol{\sigma} \mathbf{k})^{-1}$
is the bare Green function
and the dressed Green function can be written in terms of the self-energy as follows
\begin{eqnarray} \label{eq:Gret}
G(\varepsilon,\mathbf{k})=
\frac{\varepsilon-\Sigma(\varepsilon,k)+ v_0 \boldsymbol{\sigma} \mathbf{k}}%
{[\varepsilon-\Sigma(\varepsilon,k)]^2- v_0^2 k^2}.
\end{eqnarray}
The  Green function averaged over disorder within the SCBA is
shown schematically in Fig.~\ref{fig:SCBA}.
One takes into account only the one-loop diagram contributing to
the self-energy with the bare Green function replaced by the dressed one.
The self consistent equation for this self energy is simplified by its independence  on
the external momenta.
It is easy to see that one can treat different types of  SR and LR correlated
disorder on the same footing by introducing the
effective couplings $\alpha=\alpha_0+\alpha_x+\alpha_y+\alpha_z$
and $\beta=\beta_0+\beta_x+\beta_y+\beta_z$.
The one-loop self-energy diagram with the dressed Green function is given
by the integral
\begin{align}
\Sigma(\varepsilon)&=\int_\mathbf{k} 2\pi v_0^2 g(k) G_0(\varepsilon,\mathbf{k})
\nonumber \\
& = X(\varepsilon)
\int\limits_0^{\Delta/v_0} \frac{ v_0^2 g(k) k dk}{X^2(\varepsilon)-v_0^2k^2},
\label{eq:self}
\end{align}
where we have introduced the UV momentum cutoff  $\Delta/v_0$.
The function
$X(\varepsilon)=\varepsilon- \Sigma(\varepsilon)$ has to be determined
self-consistently. Once Eq.~(\ref{eq:self}) is solved the density of states
can be computed using the retarded Green function~(\ref{eq:Gret}) as follows
\begin{align}
\rho(\varepsilon)&=-\frac1{\pi} \mathrm{Im\ Tr} \int_\mathbf{k}
G^R(\varepsilon,\mathbf{k})\nonumber \\
&  = \frac1{2\pi^2 v_0^2}  \mathrm{Im} X(\varepsilon)\ln \left[-
\frac{\Delta^2}{X^2(\varepsilon)}\right]. \label{eq:DOS}
\end{align}
We now consider separately the cases of the SR and LR correlated disorder.

\textit{SR correlated disorder.}
In this case the disorder correlator reads from eq.~(\ref{eq:g-mu}) $g(k)=\alpha$
so that the SCBA equation~(\ref{eq:self}) reduces to
\begin{eqnarray}\label{eq:Xeps0}
X(\varepsilon)=\varepsilon+\frac{\alpha}2 X(\varepsilon) \ln \left[-
\frac{\Delta^2}{X^2(\varepsilon)}\right].
\end{eqnarray}
The solution of Eq.~(\ref{eq:Xeps0}) has two branches which correspond
to the retarded and advanced Green functions. They can be written explicitly
in terms of the Lambert function\cite{olver2010} $W(x)$ as follows\cite{fukuzawa09}
\begin{eqnarray} \label{eq:Xeps1}
X(\varepsilon)=\varepsilon/(\alpha W(\pm i\varepsilon/(\alpha\Gamma_0))).
\end{eqnarray}
Here and below, the upper sign corresponds to retarded and the lower sign to advanced
functions. The weak disorder introduces a new exponentially small energy scale
$\Gamma_0=\Delta e^{-1/\alpha}$. For $\varepsilon \ll \Gamma_0$ one can expand
Eq.(\ref{eq:Xeps1}) in small $\varepsilon$, this yields
$X(0)= \mp i \Gamma_0 + \varepsilon/\alpha $.
At the Dirac point  the self-energies are pure imaginary $\Sigma(0)= \mp i \Gamma_0$.
This would imply a finite
density of states at the Dirac cone that contradicts with the results obtained
by different methods. For instance, for uncorrelated   random
gauge disorder one expects $\rho(\varepsilon)=\varepsilon^{2/z-1}$ with
$z=1+\alpha$ in weak disorder case\cite{ludwig94} ($\alpha<2$)  and
$z = (8\alpha)^{1/2} -1$ in strong disorder case\cite{horovitz02} ($\alpha>2$).
It was argued in Ref.~\onlinecite{nersesyan95}  that the failure of SCBA in
the vicinity of the Dirac cone for $\varepsilon< \Gamma_0$
is due to importance of the diagrams with crossed disorder lines, neglected within
the SCBA which takes into account only the non-crossed ones. Another reason
for the failure of the SCBA is   the divergence of the fermion wavelengths
 at the Dirac point  rendering the SCBA uncontrollable, i.e.
without a small parameter. Nevertheless one can
still rely on the SCBA for energies $\varepsilon\gg \Gamma_0$.
For $\Gamma_0 \ll \varepsilon \ll \Delta$ and $\alpha\ll 1$ one can derive an
approximate solution of (\ref{eq:Xeps0}) by iteration of the Eq.~(\ref{eq:Xeps0}).
To lowest order one obtains\cite{ostrovsky06}
\begin{eqnarray}
X(\varepsilon) = \varepsilon\left (1+\alpha \ln \frac{\Delta}{\varepsilon}\right)\pm
\frac{i}2 \pi \alpha \varepsilon\left[1+2\alpha \ln \frac{\Delta}{\varepsilon}
\right].
\end{eqnarray}
Reexpressing $X \ln(-\Delta^2/X^2) $ in Eq.~(\ref{eq:DOS}) using Eq.~(\ref{eq:Xeps0})
we get the density of states
\begin{eqnarray}
\rho^{\mathrm{SR}}_{\mathrm{SCBA}}(\varepsilon)=
\frac{\varepsilon}{\pi^2 v_0^2 \alpha^2}  \mathrm{Im}\, \frac{1}{
 W(\varepsilon/(i\alpha\Gamma_0)))} .
\end{eqnarray}
For $\varepsilon\gg \Gamma_0$ this simplifies to
\begin{eqnarray}
\rho^{\mathrm{SR}}_{\mathrm{SCBA}}(\varepsilon)=
\frac{\varepsilon}{2\pi v_0^2 }\left[1+2\alpha \ln \frac{\Delta}{\varepsilon}
\right], \label{eq:DOS2}
\end{eqnarray}
while below the energy scale $\Gamma_0$ the density of states saturates at a
finite value which is in fact correct only for random potential
disorder.\cite{ostrovsky06}

\textit{LR correlated disorder.}
The disorder correlator having the form $g(k)=\beta k^{a-2}$
(see eq.~(\ref{eq:g-mu})) yields the SCBA equation of the following form
\begin{eqnarray} \label{eq:Xeps4}
X(\varepsilon)=\varepsilon-\frac{\beta \Delta^a v_0^{2-a}}{a\, X(\varepsilon)} \,
{}_2F_1\left(1,\frac{a}2,1+\frac{a}2; \frac{\Delta^2}{X^2(\varepsilon)}\right),
\end{eqnarray}
where ${}_2F_1(a,b,c;z)$ is the Gauss hypergeometric
function.\cite{olver2010}
In the limit $a\to 2$ we recover Eq.~(\ref{eq:Xeps0}) with $\alpha$ replaced by
$\beta$.
Similarly to the  previous case Eq.~(\ref{eq:Xeps4})  has two solutions
corresponding to the retarded and advanced Green functions.  At the Dirac point
the self-energies are pure imaginary: $\Sigma(0)= \mp i \Gamma_{2-a}$ and disorder
induces a generalized small energy scale
\begin{eqnarray}
&&\!\!\!\!\! \Gamma_{2-a} =  \Delta   \left(\frac{2\sin
   \left(\frac{\pi  a}{2}\right) \left(2-a +\bar{\beta}
   \right)}{(2-a)\pi \bar{\beta} }\right)^{-\frac{1}{2-a}} \nonumber \\
   && \!\!\!\!\! \approx \Delta\, e^{-1/\bar{\beta} }
     \left[1 + (2-a) \left(\frac1{2\bar{\beta}^2}+
    \frac{\pi ^2}{24}\right)+O((2-a)^2)
    \right],  \ \ \ \ \
\end{eqnarray}
where we have introduced the dimensionless disorder strength
$\bar{\beta}={\beta}(\Delta/v_0)^{a-2}$ and in the second line we have performed
an expansion to the first order in $2-a$. For finite
$2-a$ the energy scale $\Gamma_{2-a}$ is only algebraically small in disorder
at variance with the exponentially small energy scale $\Gamma_{0}$ for uncorrelated
disorder.
Iterating Eq.~(\ref{eq:Xeps4}) one obtains an
approximate solution, valid for
$\Gamma_{2-a} \ll \varepsilon \ll \Delta$ and $\beta\ll 1$.
To lowest order that solution reads
\begin{eqnarray}
X(\varepsilon)& =& \varepsilon\left (1+ \bar{\beta} U(\varepsilon) \right)\pm
\frac{i}2 \pi \bar{\beta} \varepsilon
\left[1+2 \bar{\beta} U(\varepsilon)\right], \label{eq:Xeps5} \\
U(\varepsilon)&=& -\frac{\pi}2 \left(
\frac{\Delta}{\varepsilon}\right)^{2-a}\cot\frac{\pi a}{2}-\frac1{2-a},
\end{eqnarray}
In the limit of $a \to 2$ one obtains $U(\varepsilon)\simeq \ln (\Delta/\varepsilon)$.
Substituting the solution~(\ref{eq:Xeps5}) in Eq.~(\ref{eq:DOS})
we obtain the density of states  for $\Gamma_{2-a} \ll \varepsilon$
\begin{eqnarray}
\rho^{\mathrm{LR}}_{\mathrm{SCBA}}(\varepsilon)=
\frac{\varepsilon}{2\pi v_0^2 }\left[1+\bar{\beta}\left(
\ln \frac{\Delta}{\varepsilon}+U(\varepsilon)\right)
\right],
\end{eqnarray}
which reproduces the density of states~(\ref{eq:DOS2}) in the limit
of SR correlated disorder $a \to 2$.

\section{Full counting statistics}
\label{sec:full}

\subsection{Matrix Green function formalism}

\begin{figure}
\includegraphics[width=7cm]{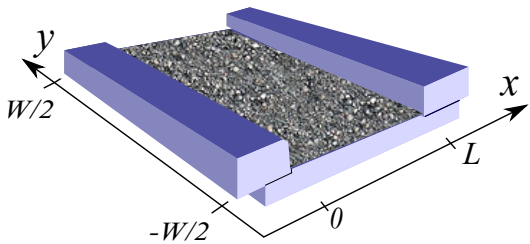}
\caption{(Color online) The setup for two-terminal transport measurements:
 a 2D disordered sample of size $W \times L$ with  perfect leads attached on opposite sides. }
  \label{fig:setup}
\end{figure}

We now consider the transport properties of
2D Dirac fermions propagating in a rectangular sample
of size $L\times W$. The schematic setup is shown in Fig.~\ref{fig:setup}.
We assume that the perfect metallic leads are  attached
to the two sides of the width $W$ with the distance $L\ll W$ between them.
We model the leads as heavily doped regions described by the
same Dirac Hamiltonian but with the
chemical potential  $\epsilon_F\gg \epsilon $ shifted far
from the chemical potential $\epsilon\simeq 0$ in the bulk which is close to the
Dirac point.  The large number of propagating
modes in the leads
are labeled by the momentum $p_n=2\pi n/W$ in $y$ direction with
$n = 0,\pm 1,.. \pm  \epsilon_F W / (2\pi v_0)$. In the limit $W\gg L$
in which many  modes $N\gg1$ contribute to transport  one
can neglect the boundary conditions at $y=\pm W/2$ and treat $p_n$ as a
continuous variable $p$. It is convenient to switch from the coordinate representation
to the mixed channel-coordinate  representation $\psi (x,y) \to \psi_n(x)$ where
$n$ enumerates the transverse modes.
Using this basis one can describe the wave functions in the leads
by two vectors
$c^{\mathrm{in}}=[\{a_n^+\},\{b_n^-\}]$ and $c^{\mathrm{out}}=[\{a_n^-\},\{b_n^+\}]$
where $a_n$ and $b_n$  refer to the amplitude of waves in the left and
in the right lead, respectively. The sign "+" refers to the waves moving to
right and the sign "-" to the waves moving to left. These two vectors are related
by the scattering matrix $\mathcal{S}$ as
$c^{\mathrm{out}}=\mathcal{S} c^{\mathrm{in}}$. In the lead subspace it has the
standard
structure:\cite{beenakker97}
\begin{eqnarray}
 \mathcal{S} =
\left(
 \begin{array}{cc}
    \hat{r} & \hat{t}' \\
    \hat{t} & \hat{r}' \\
  \end{array}
\right),
\end{eqnarray}
where we use the "hat" notation for matrices
defined in the channel space.
The conservation of particles implies that $\mathcal{S}$
is a unitary matrix and that the
four Hermitian matrices $\hat{t} \hat{t}^{\dagger}$,
$\hat{t}^{\prime \dagger}\hat{t}^{\prime \dagger}$, $1-\hat{r} \hat{r}^{\dagger}$,
$1-\hat{r}^{\prime \dagger}\hat{r}^{\prime \dagger}$
have the same set of eigenvalues $T_n$ each
of them is a real number between 0 and 1.
The transport statistics is completely determined by the matrix of transmission
amplitudes $t_{mn}$ between channels $m$ and $n$ in the leads since the transmission
probabilities of the system are given by the eigenvalues $T_n$ of the matrix
$\hat{t}^{\dagger} \hat{t}$.\cite{beenakker97}

The scattering matrix $\mathcal{S}$ relates incoming to outgoing
states. An alternative formulation is based on the  the transfer matrix
$\mathcal{T}$  which relates the states in the left lead
to states in the right lead:
$c^{\mathrm{right}}= \mathcal{T} c^{\mathrm{left}} $.
The waves in the left and the right
leads are given by the vectors $c^{\mathrm{left}}=[\{a_n^+\},\{a_n^-\}]$ and
$c^{\mathrm{right}}=[\{b_n^+\},\{b_n^-\}]$, respectively.
One can show that the eigenvalues of $\mathcal{T T}^{\dagger}$ appear in pairs
of the form $e^{\pm 2\lambda_n}$ with
$\lambda_n\ge0$ related to the transmission eigenvalues by
$ T_n=1/\cosh^2 \lambda_n$.

The transmission eigenvalues allow one to calculate a variety of
transport properties. In the limit of large number of channels one
can introduce the distribution function $P(T)$. By definition
$\int dT P(T) $ gives the total number of open channels.
The first two moments of the distribution give the Landauer conductance
\begin{equation}
G = \frac{e^2}{h} \textrm{Tr}\, \hat{t}^{\dagger} \hat{t}
= \frac{e^2}{h} \int_0^{\infty} dT\, T P(T). \label{eq:G-landauer}
\end{equation}
and the Fano factor
\begin{eqnarray}
F &=& 1- \frac{\mathrm{Tr}\,( \hat{t}^{\dagger} \hat{t})^2}{
\textrm{Tr}\, \hat{t}^{\dagger} \hat{t}} \nonumber \\
&=& 1- \left(\int_0^{\infty} dT T^2 P(T)\right)\Big{/}
\left(\int_0^{\infty} dT T P(T)\right), \ \ \ \
\end{eqnarray}
which describes the  power spectrum of
the noise due to discreetness of the charge carriers at zero frequency
and average current $I$: $P_0=2eF I$.
Note, that for graphene one has to multiply Eq.~(\ref{eq:G-landauer}) by the factor
of 4 accounting for the spin and valley degeneracy.
It is also convenient introduce the probability density $\mathcal{P}(\lambda)$
of the parameter $\lambda$ defined by  $T=1/\cosh^{2}(\lambda)$, which is naturally
 completely equivalent to $P(T)$.

In general one can write down an integral equation for the transfer matrix
with a kernel which depends on a particular realization of disorder. Iterating
the integral equation and averaging over disorder one can compute the transfer matrix
as an expansion in small disorder. However, in the case
of LR correlated disorder the forthcoming problem of computing the transmission
eigenvalues seems to be a formidable task.  Fortunately, there is an alternative
way which allows one to relate $P(T)$ to the free energy of an auxiliary field
theory. This method is based on the matrix Green function formalism
introduced by Nazarov.\cite{nazarov94} Instead of $P(T)$ the statistics
of the transmission eigenvalues can be encoded in the generating function
\begin{eqnarray}
\mathcal{F}(z)=\sum\limits_{n=1}^{\infty} z^{n-1} \mathrm{Tr}
( \hat{t}^{\dagger} \hat{t})^n= \mathrm{Tr}[\hat{t}^{-1}
\hat{t}^{\dagger\,-1}-z]^{-1}.
\end{eqnarray}
All moments of $P(T)$ can be computed using the series expansion
of $\mathcal{F}(z)$ at $z=0$. The function $\mathcal{F}(z)$ is regular
in vicinity of $z=0$ and has a brunch cut along the real axis going
from 1 to $\infty$.
Both functions are related by the Riemann-Hilbert equation
\begin{equation}
\mathcal{F}(z)=\int_0^1 \frac{P(T) dT}{T^{-1} -z}
\end{equation}
and its solution is given by the jump of $\mathcal{F}(z)$ across
the brunch cut
\begin{equation}
P(T)=\frac1{2\pi i T^2} [\mathcal{F}(1/T +i 0)-\mathcal{F}(1/T-i0)].
\end{equation}
To calculate $\mathcal{F}(z)$ we now  adopt the  matrix Green functions approach
originally developed in Ref.~\onlinecite{nazarov94} and applied to
Dirac fermions in graphene with uncorrelated
disorder in Ref.~\onlinecite{schuessler10}.
The coefficients of the series expansion of the generating function at $z=0$
can be expressed in terms of the Green functions of the system  as
\begin{eqnarray}
\mathrm{Tr} ( \hat{t}^{\dagger} \hat{t})^n= \mathrm{Tr}[\hat{v}\hat{G}^A(x,x')
\hat{v}\hat{G}^R(x',x)]^n_{x=0,x'=L}.
\end{eqnarray}
Here $\hat{v}_x= \sigma_x \hat{\mathds{1}}$ is the velocity
operator and the retarded and advanced Green functions in
the channel-coordinate representation read
\begin{eqnarray}
(\epsilon- \hat{H} \pm i0)\hat{G}^{R,A}(x,x')=\delta(x-x')\hat{\mathds{1}}.
\end{eqnarray}
The generating function can be written
as a trace of an auxiliary two component Green function defined in the
retarded-advanced (RA) space in the presence of fictitious field $z$.\cite{nazarov94}
The matrix
Green function is given by
\begin{eqnarray}  \label{eq:G-Nzarov-1}
\check{K}(x)  \check{G}(x,x')= \delta(x-x')\check{\mathds{1}},
\end{eqnarray}
where we use "check" notation for the objects defined in RA space
and the operator $\check{K}$ reads
\begin{eqnarray}
 \check{K}(x) = \left(
   \begin{array}{cc}
     \epsilon - \hat{H}  + i0 & - \sqrt{z} \hat{v} \delta(x)  \\[3mm]
     - \sqrt{z} \hat{v} \delta(x-L)   & \epsilon - \hat{H}  - i0 \\
   \end{array}
 \right).
 \label{eq:G-Nzarov-2}
\end{eqnarray}
Considering the field $z$ as a small perturbation we can rewrite
Eq.~(\ref{eq:G-Nzarov-1}) in an integral form as follows
\begin{eqnarray} \label{eq:int-eq}
\check{G}(x,x')&=&\check{G}_0(x,x') \nonumber \\
\!\!\!\! && + \sqrt{z}\int\, dx_1 \check{G}_0(x,x_1)
\check{\mathcal{V}}_{x_1} \check{G}(x_1,x') \ \ \ \ \
\end{eqnarray}
with the kernel and the inhomogeneity given by
\begin{eqnarray}
 \check{G}_0 = \left(
   \begin{array}{cc}
     \hat{G}^{R} & 0  \\[3mm]
     0   & \hat{G}^{A} \\
   \end{array}
 \right), \
  \check{\mathcal{V}}_{x} = \left(
   \begin{array}{cc}
     0 & \hat{v}\delta(x)  \\[3mm]
     \hat{v} \delta(L-x)   & 0 \\
   \end{array}
 \right).
\end{eqnarray}
One can then compute the generating function using
\begin{eqnarray} \label{eq:F-1}
\mathcal{F}(z) =\frac1{2\sqrt{z}} \int dx \,
\mathrm{Tr}[\check{\mathcal{V}}_{x} \check{G}(x,x)]
\end{eqnarray}
that can be checked by iterating Eq.~(\ref{eq:int-eq}) and substituting
in Eq.~(\ref{eq:F-1}).
One can relate $\mathcal{F}(z)$
to the object which plays the role of the free energy in the corresponding field
theory. Let us rewrite the Green function (\ref{eq:G-Nzarov-1})
in coordinate representation
using a functional integral over Grassmann variables $\bar{\psi}$ and $\psi$
\begin{eqnarray}
\check{G}(\mathbf{r},\mathbf{r}')=\frac{1}{Z}
\int \mathcal{D}\bar{\psi}\mathcal{D}\psi\, \bar{\psi}(\mathbf{r})
 \psi(\mathbf{r}')\, e^{- S}
\end{eqnarray}
with the bilinear in $\bar{\psi}$ and $\psi$ action
\begin{eqnarray}
S=\int d^2\mathbf{r}[\bar{\psi}(\mathbf{r}) \check{K} \psi(\mathbf{r})].
\label{eq:S-1}
\end{eqnarray}
The corresponding partition function and the free energy
can be written as follows
\begin{eqnarray}
Z(z)=\mathrm{Det}\,K =\int \mathcal{D}\bar{\psi}\mathcal{D}\psi\, e^{- S}, \ \ \
\Omega(z)=\ln Z.
\end{eqnarray}
It is convenient to rewrite the free energy in terms of the angle $\phi$
defined by $z=\sin^2(\phi/2)$.
Direct inspection of Eq.~(\ref{eq:int-eq}) shows that
\begin{equation}
\mathcal{F}(z) = \frac{\partial \Omega(z)}{\partial z}
=\left.\frac2{\sin \phi} \frac{\partial
\Omega (\phi)}{\partial \phi} \right|_{\phi=2\arcsin \sqrt{z}}.
\end{equation}
The distribution of transmission eigenvalues $\mathcal{P}(\lambda)$
can be calculated using the following relation
\begin{eqnarray}
\mathcal{P}(\lambda)= \left.\frac2{\pi}\mathrm{Re} \frac{\partial
\Omega (\phi)}{\partial \phi} \right|_{\phi=\pi +2i\lambda },
\end{eqnarray}
from which one can easily derive $P(T)$. Therefore, we have four
equivalent descriptions of the transport properties in terms of one
of the following functions: $P(T)$, $\mathcal{P}(\lambda)$, $\mathcal{F}(z)$ or
$\Omega(\phi)$. Any of these functions can be used for computing
the conductance or the Fano factor. For instance, using the free energy
one can derive the expressions
\begin{eqnarray} \label{eq:GF-0}
G=\frac{2e^2}{h}\Omega''(0), \ \ \ F=\frac13-\frac23 \frac{\Omega^{(\mathrm{{IV}})}
(0)}{\Omega''(0)},
\end{eqnarray}
where the derivatives are taken at $\phi=0$. Note that in the case of graphene
the conductance (\ref{eq:GF-0}) is given as expected per Dirac species, {\it i.e.} per spin and per valley.

\subsection{Expansion in disorder and diagrammatics }
In what follows we restrict our consideration to transport around the Dirac
cone $\epsilon=0$.
The action (\ref{eq:S-1}) for the system including the metallic leads
can be calculated using the kernel~(\ref{eq:G-Nzarov-2}) with
Hamiltonian~(\ref{eq:Dirac1}) in which the free part is modified to
\begin{eqnarray} \label{eq:H-1}
H_0 = -\mu(x) - i v_0 (\sigma_x\partial_x+\sigma_y\partial_y).
\end{eqnarray}
Here $\mu(x)=0$ for  $0<x<L$ and $+\infty$ otherwise accounts for the leads
with very high chemical potential.
Above we have treated the auxiliary field $z$ as  a perturbation.
Here we split the kernel~(\ref{eq:G-Nzarov-2})
into the free part including the auxiliary field and the interaction part:
$\check{K}=\check{K}_0+\check{K}_V$, where $\check{K}_0$ is computed using
Eq.~(\ref{eq:H-1}) and $\check{K}_V$ is diagonal in RA space.

\begin{figure}
\includegraphics[width=80mm]{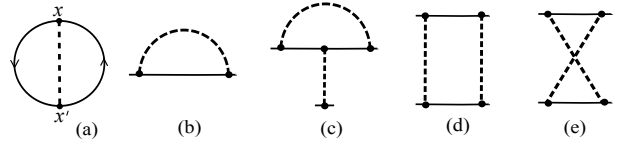}
\caption{The one-loop diagrams contributing to (a) free energy;
(b) propagator; (c)-(e) disorder renormalization. }
  \label{fig:free-energy}
\end{figure}

We are now in the position to average the free energy over disorder.
To that end we use the replica trick and introduce
$n$ copies of the original system. Performing averaging over disorder
we obtain the replicated action in the following form
\begin{eqnarray}
S& =& \sum\limits_{\alpha=1}^n
\int d^2 \mathbf{r} \bar{\psi}_{\alpha}(\mathbf{r}) \check{K}_0
\psi_{\alpha}(\mathbf{r})
\nonumber \\
&&  + \pi v_0^2   \sum\limits_{\alpha,\beta=1}^n \sum\limits_{\mu=0}^3
\int d^2 \mathbf{r} \int d^2 \mathbf{r}' g_{\mu}(\mathbf{r}-\mathbf{r}')
 \nonumber \\
&& \times  [\bar{\psi}_{\alpha}(\mathbf{r})\Sigma_{\mu} \psi_{\alpha}(\mathbf{r})]
[\bar{\psi}_{\beta}(\mathbf{r})\Sigma_{\mu} \psi_{\beta}(\mathbf{r})]
\end{eqnarray}
where $\check{K}_0$ is given by Eq.~(\ref{eq:G-Nzarov-2}) with
Hamiltonian (\ref{eq:H-1}) and we defined the matrix
$\Sigma_{\mu} = \check{\mathds{1}}\otimes\sigma_{\mu}$. In what follows
we set $v_0=1$ unless it is written explicitly.
The bare Green function corresponding to $\check{K}_0$
 in Eq.~(\ref{eq:G-Nzarov-1}) can be written in coordinate
representation as
\begin{eqnarray}
&&G_0(x,x';y)=\frac1{4L \cos (\phi/2)} \times \nonumber \\
&& \left(
\begin{array}{cccc}
  \frac{\cos \phi (\frac{1}{2} - x^0_0 )}{ i \sin \pi x^0_0}  &
  \frac{\cos \phi (\frac{1}{2} - x^1_0 )}{ i \sin \pi x^1_0}  &
  \frac{\sin  \phi(1-x^0_0 )}{\sin \pi x^0_0 } &
  \frac{\sin \phi x^1_0 }{\sin \pi x^1_0 } \\
  \frac{\cos \phi (\frac{1}{2} - x^1_1 )}{ i \sin \pi x^1_1}  &
  \frac{\cos \phi (\frac{1}{2} - x^0_1 )}{ i \sin \pi x^0_1}  &
  \frac{\sin  \phi x^1_1 }{\sin \pi x^1_1 } &
  \frac{\sin \phi(1-x^0_1 )}{\sin \pi x^0_1 } \\
   \frac{\sin  \phi x^0_0 }{\sin \pi x^0_0 } &
   \frac{\sin  \phi x^1_0 }{\sin \pi x^1_0 } &
 \frac{i \cos \phi (\frac{1}{2} - x^0_0 )}{  \sin \pi x^0_0}  &
  \frac{\cos \phi (\frac{1}{2} + x^1_0 )}{ i \sin \pi x^1_0}  \\
    \frac{\sin  \phi x^1_1 }{\sin \pi x^1_1 } &
   \frac{\sin  \phi x^0_1 }{\sin \pi x^0_1 } &
  \frac{\cos \phi (\frac{1}{2} + x^1_1 )}{ i \sin \pi x^1_1}  &
  \frac{i \cos \phi (\frac{1}{2} - x^0_1 )}{ \sin \pi x^0_1}
\end{array}
\right), \nonumber \\ \label{eq:G-Naz-1}
\end{eqnarray}
where we have introduced the shorthand notation
$x^{k}_{l}= (x + (-1)^k x' +(-1)^l i y)/2L$.

We will first reproduce the known results for the free Dirac fermions
i.e. for clean graphene. To that end we rewrite the free energy (\ref{eq:F-1})
in the coordinate representation as a function of $\phi$:
\begin{eqnarray}
\mathcal{F}(\phi)= \frac{W}{2 \sin \phi/2} \int dx\, \mathrm{Tr}
\left[ \check{\mathcal{V}}_x \check{G}_0(x,x,0)\right].
\end{eqnarray}
Substituting the bare Green function~(\ref{eq:G-Naz-1}) we obtain
the generating function and the corresponding free energy
\begin{eqnarray}
\mathcal{F}_0(z)=\frac{W \arcsin\sqrt z }{\pi L  \sqrt{z-z^2}}, \ \ \ \
\Omega_0(\phi)= \frac{W \phi^2}{4\pi L}. \label{eq:free-energy-0}
\end{eqnarray}
The corresponding distribution of transmission eigenvalues then reads
\begin{eqnarray} \label{eq:dorokhov}
P_0(T)=\frac{W}{2\pi L}\frac1{T\sqrt{1-T}}.
\end{eqnarray}
The distribution $P(T)$ is expected to be integrable and the integral
gives the total number of open channels. However the integral of
Eq. (\ref{eq:dorokhov}) diverges logarithmical at $T=0$: we
need to introduce a cutoff $T_{\mathrm{min}}\sim e^{-2\epsilon L/v_0}$. However,
the contribution of this cutoff to the higher moments is found to be exponentially small.
Equation~(\ref{eq:dorokhov}) coincides with the well-known result obtained by
Dorokhov for the disordered metallic wires.\cite{dorokhov83}
Thus, the transport of clean 2D Dirac
fermions resembles the diffusive transport of non-relativistic electrons in
quasi-one-dimensional systems in the presence of disorder. This  nontrivial
result can be explained by existence of the evanescent modes.

\subsection{The lowest order correction to free energy}

The perturbative corrections to the free energy of the system due
to disorder can be expressed as a sum of loop diagrams without external
legs. The lowest order contributions to the free energy
are given by the one loop diagrams schematically shown in
Fig.~\ref{fig:free-energy}(a).  The solid lines denote the propagator
(\ref{eq:G-Naz-1}) in the presence of the boundary auxiliary field $\phi$ and
the dashed lines stands for disorder correlation functions.  There are six
topologically  equivalent diagrams with different disorder correlators corresponding
to SR and LR correlated disorder of three types: random potential, random gauge (with
two components $x$ and $y$) and random mass. The corresponding integrals
have the form
\begin{equation} \label{eq:diag-naz}
\Omega_{\mathrm{SR,LR}} = \pi \int d^2\mathbf{r} d^2\mathbf{r}'
g_{\mu}(\mathbf{r}-\mathbf{r}')
\Tr[\check{\Sigma}_{\mu} \check{G}(\mathbf{r},\mathbf{r}')
 \check{\Sigma}_{\mu} \check{G}(\mathbf{r}',\mathbf{r})].
\end{equation}
with $g_{\mu}(\mathbf{r})$, $\mu=0, x,y,z$ given by Eq.~(\ref{eq:g-mu})
where we retain only the SR or LR part. These integrals diverge for
$\mathbf{r}\to \mathbf{r}'$,  however, the divergent terms turn out to be
$\phi$-independent, and thus do not contribute to the physical quantities
(\ref{eq:GF-0}). Hence, it is convenient
to consider the first derivative of the free energy with respect to $\phi$ which
is finite and determines the physical observables.
The $\phi$-dependent parts of diagrams with dashed line corresponding to the
three types of SR correlated disorder were computed in Ref.~\onlinecite{schuessler10}.
The result giving the linear in $\alpha_{\mu}$ correction to the free energy of
the clean sample (\ref{eq:free-energy-0}) reads
\begin{eqnarray}
\Omega'_{\mathrm{SR}}(\phi) = \frac{W \phi}{2\pi L}
\left(\alpha_0-\alpha_z \right). \label{eq:free-energy-002}
\end{eqnarray}

Note that the SR  correlated random gauge potential does
not contribute to the
transport properties to one-loop order and this holds also to two-loop order.
This is in agreement with the arguments of
Ref.~\onlinecite{schuessler10} that at zero energy the
gauge potential can be eliminated by a pseudogauge transformation of
the wave function. As a result,  the transport properties are not influenced
by random gauge potential despite the fact that it
gives rise to a multifractal wave function $\Psi(\mathbf{r})$ with a
disorder strength dependent spectrum of multifractal
exponents.\cite{ludwig94,chamon96,castillo97,comtet98,carpentier01}
\begin{figure}
\includegraphics[width=80mm]{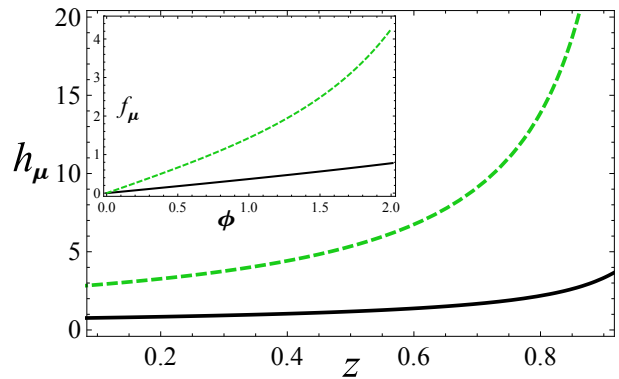}
\caption{(Color online) The one-loop LR disorder corrections to
the generating function of transmissions~(\ref{eq:gen-fun-2}) for $a=1$:
$h_0(z)$ - solid black line and
 $h_z(z)$ - dashed green line.
 \textbf{Inset}. The one-loop LR  disorder corrections to the free energy
(\ref{eq:free-energy-02}) for  $a=1$:
$f_0(\phi)$ - solid black line and
 $f_z(\phi)$ - dashed green line.}
  \label{fig:f-functions}
\end{figure}

The three diagrams with dashed line associated with correlation functions of
the LR correlated disorder are computed in Appendix~\ref{sec:integrals}.
The corresponding corrections to the free energy read
\begin{eqnarray}
\Omega'_{\mathrm{LR}}(\phi)  &=& \frac{W }{2\pi L^{a-1}}
 \left[f_0(\phi) \beta_0-f_z(\phi) \beta_z
\right]. \nonumber \\
&& \label{eq:free-energy-02}
\end{eqnarray}
We found that the LR correlated random gauge potential does
not contribute to the transport properties to lowest order in disorder strength.
The functions $f_{\mu}(\phi)$
are given by the following double integrals:
\begin{eqnarray}
&& f_{0,z}(\phi)= \int \limits_{0}^{\infty}dy  \int\limits_0^{\pi} d c
 \frac{4 \pi ^{a-2} \mathcal{A}_a y \sinh (y \phi/\pi ) }{\left(c^2+y^2\right)^{a/2}}
     \nonumber\\
&& \times \left\{\pm \frac{  1}{ \sinh y}
    \right.
  \left[\arctan\left( \frac{1-\cos c \cosh   y
  }{\sin c \sinh y }\right) -\frac{\pi}{2}\right] \nonumber \\
 &&   \left.  + \frac{\pi -c}{\cosh y - \cos c  }  \right\},\label{eq:fz}
\end{eqnarray}
where the upper sign corresponds to $f_0(\phi)$ and the lower sign to $f_z(\phi)$.
The functions $f_{\mu}(\phi)$ computed numerically for a particular value
of $a=1$ are shown in inset of Fig.~\ref{fig:f-functions}.

\begin{figure}
\includegraphics[width=80mm]{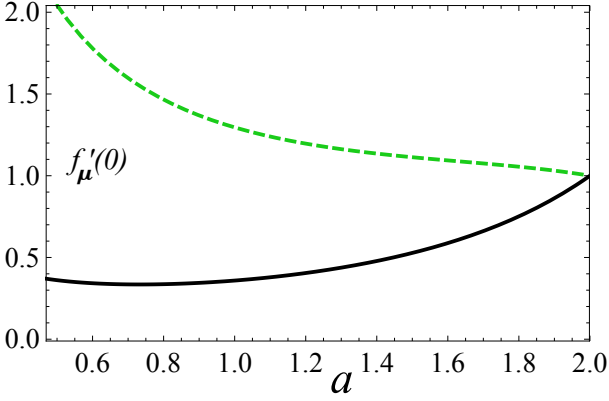}
\caption{(Color online) The one-loop LR disorder corrections to
the conductance~(\ref{eq:condactance-2}) as functions of $a$:
$f_0'(0)$ - solid black
line and  $f_z'(0)$ - dashed green line.}
  \label{fig:condactance}
\end{figure}

It is known that even in the case of uncorrelated disorder the
lowest order corrections to the density of states and the transport properties
of 2D Dirac fermions are insufficient. The leading corrections
can be summed up with the help of the renormalization group methods that
will be done in the next section.

\section{Weak disorder renormalization group}
\label{sec:RG}

Straightforward  dimensional analysis shows that the SR correlated disorder
is dimensionless
and hence marginally relevant in $d=2$. The LR correlated disorder is
relevant in $d=2$ for $a<2$. In what follows it is convenient to introduce the
rescaled disorder strengths: $\alpha(L)=\tilde{\alpha}(L)$ and
$\beta(L)=\tilde{\beta}(L)L^{a-2}$.
The lowest order corrections to the disorder strength and energy
are given by the one-loop diagrams (b)-(e) shown in Fig.~\ref{fig:free-energy}.
The corresponding RG flow equations read
\begin{subequations}\label{eq:flow-e}
\begin{eqnarray}
\frac{\partial\tilde{\alpha}_0}{\partial \ln L } &=& 2\tilde{\alpha}_0(
\tilde{\alpha}_{0}+\tilde{\beta}_{0}+\tilde{\alpha}_{\bot}+
\tilde{\beta}_{\bot}+\tilde{\alpha}_{z}+\tilde{\beta}_{z}) \nonumber \\
&&  +2
(\tilde{\alpha}_{\bot}+\tilde{\beta}_{\bot})(\tilde{\alpha}_{z}+\tilde{\beta}_{z}),
\label{eq:flow-e1} \\
\frac{\partial\tilde{\beta}_0}{\partial \ln L } &=& (2-a)\tilde{\beta}_0 +
2\tilde{\beta}_0 ( \tilde{\alpha}_{0}+\tilde{\beta}_{0}+\tilde{\alpha}_{\bot}+\tilde{\beta}_{\bot} \nonumber \\
&& +\tilde{\alpha}_{z}+\tilde{\beta}_{z}),\\
\frac{\partial\tilde{\alpha}_{\bot}}{\partial \ln L } &=& 4(\tilde{\alpha}_{0}+\tilde{\beta}_{0})
(\tilde{\alpha}_{z}+\tilde{\beta}_{z}) , \\
\frac{\partial\tilde{\beta}_{\bot}}{\partial \ln L } &=& (2-a)\tilde{\beta}_{\bot},\\
\frac{\partial\tilde{\alpha}_z}{\partial \ln L } &=& - 2 \tilde{\alpha}_z^2-2\tilde{\alpha}_z\tilde{\beta}_z
 + 2 (\tilde{\alpha}_z+ \tilde{\alpha}_{0}+\tilde{\beta}_{0} ) (\tilde{\alpha}_{\bot}+\tilde{\beta}_{\bot}) \nonumber \\
 && - 2 \tilde{\alpha}_z (\tilde{\alpha}_{0}+\tilde{\beta}_{0}), \\
\frac{\partial\tilde{\beta}_z}{\partial \ln L } &=& (2-a)\tilde{\beta}_z - 2 \tilde{\beta}_z^2
-2\tilde{\alpha}_z\tilde{\beta}_z \nonumber \\
&& + 2 \tilde{\beta}_z (\tilde{\alpha}_{\bot}+\tilde{\beta}_{\bot}-\tilde{\alpha}_{0}-\tilde{\beta}_{0}), \\
\frac{\partial \ln \tilde{\varepsilon}}{\partial \ln L } &=& 1+
\tilde{\alpha}_{0}+\tilde{\beta}_{0}+\tilde{\alpha}_{\bot}+
\tilde{\beta}_{\bot}+\tilde{\alpha}_{z}+\tilde{\beta}_{z}, \label{eq:flow-e2}
\end{eqnarray}
\end{subequations}
where we used the notation $\tilde{\alpha}_{\bot}=\tilde{\alpha}_x+\tilde{\alpha}_y$
and $\tilde{\beta}_{\bot}=\tilde{\beta}_x+\tilde{\beta}_y$.
Note, that in deriving the flow equations (\ref{eq:flow-e})
we assume that $2-a$ is small and perform
$2-a$ expansion similar to $d-2$  expansion in higher dimensions. In general
in the presence of LR correlated disorder one has to rely on the double expansion
in $2-a$ and $d-2$ similar to that for the $\phi^4$ model with correlated random bond
disorder where one uses a double expansion in $4-a$ and $4-d$ at the upper critical
dimension.\cite{weinrib83}

The bare values of the disorder strengths and energy corresponding to the
microscopic scale provide the initial condition for the RG
equations~(\ref{eq:flow-e}).
The  renormalized disorder strengths, $\alpha(L)$, $\beta(L)$, and
the energy $\varepsilon(L)$ acquire scale dependence on the
ultraviolet cutoff length $L$.
One has to stop the renormalization
when either  $L$ reaches the system size or the energy $\tilde{\varepsilon}$
reaches the value
of the cutoff $\Delta$ or the disorder strengths become of order one.\cite{schuessler09} Once the
renormalization has been done one can compute the observables by substituting the
renormalized quantities into the results of the perturbation theory.

\begin{figure}
\includegraphics[width=80mm]{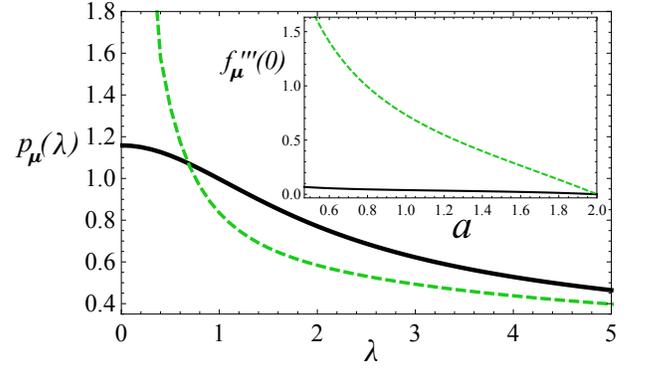}
\caption{(Color online)
 The LR correlated disorder corrections to the transmission
eigenvalue distribution $P(\lambda)$ given by Eq.~(\ref{eq:transmission-2})
for $a=1.5$: $p_0(\lambda)$ - solid black line and
 $p_z(\lambda)$ - dashed green line as a function of $\lambda$.
 \textbf{Inset.}  The LR correlated disorder corrections to the Fano factor
 given by Eq.~(\ref{eq:fano-0}):
$f_0'''(0)$ - solid black line and
 $f_z'''(0)$ - dashed green line, as functions of $a$.}
  \label{fig:fano-0}
\end{figure}

To renormalize the corrections to the free energy (\ref{eq:free-energy-002})
and (\ref{eq:free-energy-02}) we have to replace the bare coupling constants
by the renormalized ones. As a result we obtain
\begin{eqnarray}
\Omega'_{\mathrm{SR}}(\phi) &=& \frac{W \phi}{2\pi L}
\left[\tilde{\alpha}_0(L)-\tilde{\alpha}_z(L) \right],
\label{eq:free-energy-0021} \\
\tilde{\Omega}'_{\mathrm{LR}}(\phi)  &=& \frac{W }{2\pi L}
 \left[f_0(\phi) \tilde{\beta}_0(L) -f_z(\phi) \tilde{\beta}_z(L)
\right]. \ \label{eq:free-energy-2}
\end{eqnarray}
Thus, the SR correlated disorder does not modify the pseudodiffusive behavior
to lowest order ($\Omega \sim \phi^2$) and the distributions of transmission
eigenvalues is still given by the Dorokhov distribution (\ref{eq:dorokhov}).
The deviation from the pseudodiffusive regime can be found only to second order
in disorder and the corresponding two-loop corrections have the
form\cite{schuessler10}
\begin{eqnarray}
\Omega_{\mathrm{SR}}^{\mathrm{2loop}}&=&\frac{W\phi^2}{4\pi L}
[(\tilde{\alpha}_0+\tilde{\alpha}_z)^2
\omega_1(\phi)\nonumber\\
&&+(\tilde{\alpha}_0+3\tilde{\alpha}_z)
(\tilde{\alpha}_0-\tilde{\alpha}_z)] \omega_2(\phi), \label{eq:2loop}
\end{eqnarray}
where $\omega_1(\phi)=\mathrm{const}-\psi(\phi/\pi)-\psi(-\phi/\pi)$
and $\omega_2(\phi)=\mathrm{const}+\pi^2(\phi\cot \phi-1)/\phi^2$. Here $\psi(x)$
is the digamma function.\cite{olver2010}

On the contrary the LR correlated disorder
leads to deviation from pseudodiffusive transport already to lowest order in
disorder. Indeed the renormalized  corrections to the generating
function and the distribution of transmission eigenvalues read
\begin{eqnarray}
{\mathcal{F}}_{\mathrm{LR}}(z)  &=& \frac{W }{2\pi L}
 \left[h_0(z) \tilde{\beta}_0 -h_z(z) \tilde{\beta}_z
\right], \ \label{eq:gen-fun-2} \\
{\mathcal{P}}_{\mathrm{LR}}(\lambda)  &=& \frac{W }{2\pi L}
 \left[p_0(\lambda) \tilde{\beta}_0
 -p_z(\lambda) \tilde{\beta}_z
\right]. \ \label{eq:transmission-2}
\end{eqnarray}
Here $h_{\mu}(z)=2f_{\mu}(2 \arcsin \sqrt{z})/\sin\phi$ and
$p_{\mu}(\lambda)=2\mathrm{Re}f_{\mu}(\pi+2i\lambda)/\pi$.
The functions $h_{\mu}(z)$ and $p_{\mu}(\lambda)$   for particular values of $a$
are shown in Figs. (\ref{fig:f-functions}) and (\ref{fig:fano-0}).
Note that the distribution (\ref{eq:transmission-2}) can be used for direct
calculation of transmissions moments for $1<a<2$. This is different from the
two-loop correction (\ref{eq:2loop}) due to SR correlated
disorder found in Ref.~\onlinecite{schuessler10}: the corresponding contributions
to $\mathcal{P}(\lambda)$ diverge at $\lambda=0$ in a non-integrable way. This
divergence
has been attributed to the breakdown of perturbative expansion in small disorder
close to $\lambda=0$, i.e. for $T\approx 1$.
Nevertheless even for $a\le 1$ one can compute the transport characteristics directly
from the free energy.
The correction to the conductance is given by
\begin{eqnarray}
G_{\mathrm{LR}}= \frac{ e^2 W }{\pi h L}
 \left[f'_0(0) \tilde{\beta}_0  -f'_z(0) \tilde{\beta}_z
\right]. \label{eq:condactance-2}
\end{eqnarray}
and the Fano factor can be written as
\begin{eqnarray}
&& F= \frac13  - \left.
 \frac23 \frac{f'''_0 \tilde{\beta}_0-f'''_z \tilde{\beta}_z}{
 1+\alpha_0-\alpha_z+f'_0
 \tilde{\beta}_0 -f'_z \tilde{\beta}_z} \right|_{\phi=0}. \ \ \ \ \ \
 \label{eq:fano-0}
\end{eqnarray}
$f_{\mu}'(0)$ and $f_{\mu}'''(0)$ as functions of $a$ are shown in
Figs.~(\ref{fig:condactance}) and  (\ref{fig:fano-0}), respectively.
The case of generic disorder requires a numerical
solution of the flow equations which strongly depends on the particular
values of bare couplings. In the present paper we restrict our analysis to
three cases when the system
has only one type of disorder: random scalar potential, random gauge potential
or random mass disorder.

\begin{figure}
\includegraphics[width=80mm]{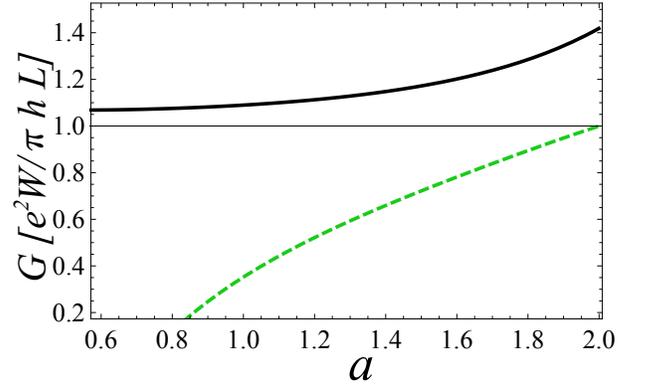}
\caption{(Color online) The  conductance
as a function of $a$ in the presence of random potential for $L/l_0=0.5$ -
solid black line and in the presence of
random mass - dashed green line.}
  \label{fig:con}
\end{figure}

\subsection{Random scalar disorder}
Let us start the discussion of the different types of disorder with random
scalar potential.
In the presence of both SR and LR correlated scalar potential the solution
of the flow equations
\begin{eqnarray}
\frac{\partial\tilde{\alpha}_0}{\partial \ln L } &=& 2\tilde{\alpha}_0(
\tilde{\alpha}_{0}+\tilde{\beta}_{0}),  \\
\frac{\partial\tilde{\beta}_0}{\partial \ln L } &=& (2-a)\tilde{\beta}_0 +
2\tilde{\beta}_0 ( \tilde{\alpha}_{0}+\tilde{\beta}_{0}), \label{eq:beta-0-flow} \\
\frac{\partial \ln \tilde{\varepsilon}}{\partial \ln L } &=& 1+
\tilde{\alpha}_{0}+\tilde{\beta}_{0}, \label{eq:eps-0-flow}
\end{eqnarray}
can be computed only numerically. However, we have found that the
LR correlated disorder dominates over SR correlated disorder at large $L$
for all bare disorder strengths such that $\tilde{\alpha}\leq\tilde{\beta}$.
Since LR correlated disorder
does not generate SR disorder itself we restrict ourselves  to the case of
pure LR disorder. Here and below we will measure the length in units of the
bare ultraviolet cutoff
given by $v_0/\Delta$.
The  solution of the flow equation~(\ref{eq:beta-0-flow})
with the initial conditions
$\tilde{\beta}_0(1)=\bar{\beta}_0=\beta(\Delta/v_0)^{a-2}$ reads
\begin{eqnarray}
\tilde{\beta}_0(L)=(a-2)\left[2+L^{a-2}(a-2(1+\bar{\beta}_0))
/\bar{\beta}_0)\right]^{-1}. \label{eq:beta0L}
\end{eqnarray}
The running disorder strength grows with the scale. At the Dirac cone the
renormalization has to be stopped either at the scale of the system size
or at the scale at which the disorder strength becomes of order unity.
This scale computed from Eq.~(\ref{eq:beta0L}) reads
\begin{eqnarray}
l_0=\left(\frac{2-a+2\bar{\beta}_0}{(4-a)\bar{\beta}_0}\right)^{1/(2-a)}.
\end{eqnarray}
$l_0$ is nothing but the zero-energy mean free path. For the system size $L<l_0$
one can rewrite the running disorder strength in terms of $l_0$ as follows
\begin{eqnarray}
\tilde{\beta}_0(L)=\frac{(2-a)}{(4-a)(l_0/L)^{2-a}-2}. \label{eq:beta0L-2}
\end{eqnarray}
For finite energy the renormalization
is limited by the scale at which the energy becomes of order of $\Delta$.
Substitution of the solution (\ref{eq:beta0L}) to the flow equation for the
energy (\ref{eq:eps-0-flow}) yields
\begin{eqnarray}
\tilde{\varepsilon}(L)=\varepsilon L \left[1-2 \bar{\beta}_0 ( L^{a-2}-1)/(a-2)
\right]^{-1/2}.
\end{eqnarray}
The renormalization stops when the running energy
$\tilde{\varepsilon}(L)$ reaches the cutoff value $\Delta$ at the scale
\begin{eqnarray}
L_{\Delta}(\varepsilon)= \frac{\Delta}{\varepsilon }
\left[1-2 \bar{\beta}_0 ( (\Delta/|\varepsilon|)^{a-2}-1)/(a-2)
\right]^{1/2}.
\end{eqnarray}
The competition between $L_{\Delta}$ and $l_0$ introduces a new exponentially
small (in the limit $a\to 2$) in disorder energy scale $\Gamma_{2-a}$ given
by equation  $L_{\Delta}(\Gamma_{2-a})=l_0$,
\begin{eqnarray}
&&\!\!\!\!\!\Gamma_{2-a} = \Delta\, l_0^{-1}\,
\sqrt{\frac{a-2-2 \bar{\beta}_0^2}{a-2-2 \bar{\beta}_0}}
\approx
\Delta e^{-\frac{1}{2 \bar{\beta}_0}+\frac{1}{2}}
{\bar{\beta}_0}^{1/2} \nonumber \\
&&\!\!\!\!\! \times \left[1+(2-a)
   \left(\frac3{8 \bar{\beta}_0^2} -
   \frac{1}{4 \bar{\beta}_0}-\frac18\right) + O((2-a)^2) \right]. \ \ \ \
\end{eqnarray}
For $\varepsilon\gg\Gamma_{2-a}$
the density of states can be found using the following scaling arguments.
The running density of states approaches $\tilde{\rho} =\Delta/(2\pi v_0^2)$
at $L=L_{\Delta}$. Taking into account that the density of states
scales as $\tilde{\rho} \tilde{\varepsilon}=\rho \varepsilon L^2$ one
can write the bare density of states as
\begin{eqnarray}
\rho(\varepsilon)=\frac{|\varepsilon|}{2\pi v_0^2}
\left[1-2 \bar{\beta}_0 ( (\Delta/|\varepsilon|)^{a-2}-1)/(a-2)
\right]^{-1}.
\end{eqnarray}
For $\varepsilon<\Gamma_{2-a}$ the density of states saturates at a finite
value. This picture is in qualitative agreement with the prediction of SCBA
computed in Sec.~\ref{sec:SCBA}. The results for SR correlated disorder case
obtained in Ref.~\onlinecite{ostrovsky06}
can be reproduced by taking the limit of $a\to2$. For instance, in this limit
we have $\Gamma_0\approx \Delta\, {\bar{\beta}_0}^{1/2} e^{-1/2\bar{\beta}_0}$.

\begin{figure}
\includegraphics[width=80mm]{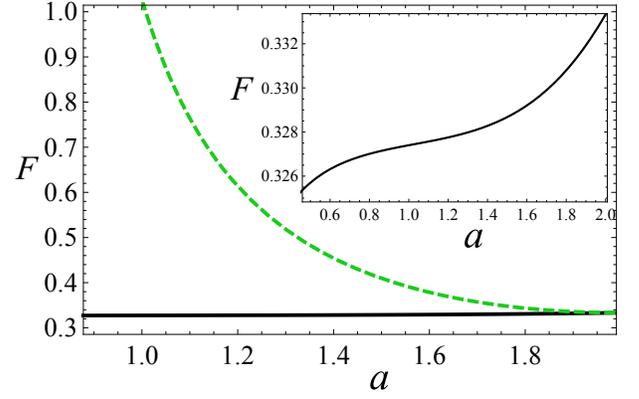}
\caption{(Color online) The Fano factor
as a function of $a$ in the presence of random potential for $L/l_0=0.5$ -
solid black line (also Inset) and random mass - dashed green line.}
  \label{fig:fano}
\end{figure}

The conductance and the Fano factor in the ballistic regime $L<l_0$ at
the Dirac cone are given by
\begin{eqnarray}
G&=&\frac{e^2}{\pi h}\frac{W}{L}\left[1+ f_0'(0)\
\tilde{\beta}_0\left(L\right) \right],\\
F&=& \frac13  -
 \frac23 \frac{f'''_0(0) \tilde{\beta}_0(L)}{1+f'_0(0) \tilde{\beta}_0(L)},
\end{eqnarray}
where $\tilde{\beta}_0(L)$ is given by Eq.~(\ref{eq:beta0L-2}). The conductance
and the Fano factor computed for $L/l_0=0.5$  are shown in Figs.~(\ref{fig:con}) and
(\ref{fig:fano}) as functions of~$a$. The correction to the conductance due
to LR correlated disorder in the ballistic
regime is positive and increases with $a$ while the correction to the Fano factor
is small and negative.

\subsection{Random gauge potential}
We now turn to the case of random gauge potential.
Inspired mostly by its relation to the
Quantum Hall transitions,\cite{ludwig94} this problem has previously motivated
numerous studies of the multifractal spectrum for critical wave
functions.\cite{chamon96,castillo97,comtet98,carpentier01} Here we
are mostly interested in the density of states and also
transport properties of such Dirac fermions with correlated random gauge
potential. In this case the flow equations reduce to
\begin{eqnarray}
\frac{\partial\tilde{\alpha}_{\bot}}{\partial \ln L } &=& 0 , \  \  \
\frac{\partial\tilde{\beta}_{\bot}}{\partial \ln L } = (2-a)\tilde{\beta}_{\bot},\\
\frac{\partial \ln \tilde{\varepsilon}}{\partial \ln L } &=& 1+
\tilde{\alpha}_{\bot}+
\tilde{\beta}_{\bot}. \label{eq:rgp-flow-energy}
\end{eqnarray}
The renormalized coupling constants have the trivial flow
$\tilde{\alpha}_{\bot}(L)=\bar{\alpha}_{\bot}$ and
$\tilde{\beta}_{\bot}(L)=\bar{\beta}_{\bot} L^{2-a}$ where
the bare disorder strengths are $\bar{\alpha}=\alpha$ and
$\bar{\beta}_{\bot}={\beta}_{\bot} (\Delta/v_0)^{a-2}$.
The LR disorder strength reaches unity at the scale
$l_0=\bar{\beta}_{\bot}^{-1/(2-a)}$.
Substituting
the running couplings to the flow equation for the energy~(\ref{eq:rgp-flow-energy})
we obtain
\begin{eqnarray}
\tilde{\varepsilon}(L)= \varepsilon L^{1+\bar{\alpha}_{\bot}}
\exp\left[\tilde{\beta}_{\bot} (L^{2-a}-1)/(2-a)\right].
\end{eqnarray}
One has to stop renormalization at the scale $L_{\Delta}$ such that
$\tilde{\varepsilon}(L_{\Delta})=\Delta$. In the case of the system with only
SR correlated random gauge disorder this scale is given by
$L_{\Delta}^{\mathrm{SR}}= \left( {\varepsilon}/{\Delta}\right)^{-1/z}$,
where we have introduced  the dynamic critical exponent $z=1+\bar{\alpha}_{\bot}$.
Note that this exponent is non-universal and depends on the strength of disorder.
In the presence of  LR correlated disorder the cutoff scale computed up to
subleading logarithmic corrections is
\begin{eqnarray}
L_{\Delta}^{\mathrm{LR}}= \left(\frac{2-a}{\bar{\beta}_{\bot}}
\ln \frac{\Delta}{\varepsilon}\right)^{1/(2-a)}.
\end{eqnarray}
However, one has to stop renormalization at $l_0$ for $l_0<L_{\Delta}^{\mathrm{LR}}$
that introduces a new energy scale $\Gamma_{2-a}=\Delta e^{-1/(2-a)}$ which is
exponentially small for $a\to 2$.
The bare density of states is then given by
$\rho=\tilde{\rho} \tilde{\varepsilon}/(\varepsilon L_{\Delta}^{2})$
with $\tilde{\rho} =\Delta/(2\pi v_0^2)$. Substituting the renormalized cutoff
scale we obtain for the SR correlated disorder a non-universal power-law
behavior:
\begin{eqnarray}\label{eq:DOS-RG-SR}
\rho_{\mathrm{SR}}(\varepsilon)&=&\frac{\Delta}{2\pi v_0^2}
\left( \frac{\varepsilon}{\Delta}\right)^{(2-z)/z}, \ \ \ z=1+\bar{\alpha}_{\bot}
\end{eqnarray}
which was firstly derived in Ref.~\onlinecite{ludwig94}.
In the case of LR random gauge disorder we have
\begin{eqnarray}
\rho_{\mathrm{LR}}(\varepsilon)&=&\frac{\Delta^2}{2\pi v_0^2} \frac1{\varepsilon}
\left(\frac{2-a}{\bar{\beta}_{\bot}} \ln \frac{\Delta}{\varepsilon}
\right)^{-2/(2-a)} \nonumber \\
&=&\frac{1}{2\pi \varepsilon}
\left(\frac{2-a}{{\beta}_{\bot}} \ln \frac{\Delta}{\varepsilon}
\right)^{-2/(2-a)}, \label{eq:DOS-gauge-1}
\end{eqnarray}
where in the last line we used the definition of the dimensionless disorder
strength so that the dependence on the ultraviolet cutoff drops out from
the density of states.
Presumably Eq.~(\ref{eq:DOS-gauge-1}) is valid only for $\varepsilon>\Gamma_{2-a}$.
In Sec.~\ref{sec:bosonization} we apply bosonisation technique
to compute the density of states down to zero energy and show that
the scaling behavior~(\ref{eq:DOS-gauge-1}) actually holds up down to zero energy.

We have found above that the LR correlated random gauge disorder
does not contribute to transport  at the Dirac cone.
There are some general arguments that any random gauge potential cannot
modify the transport properties.
Let us first briefly recall the argument of Ref.~\onlinecite{schuessler09}.
To start, we consider the Hamiltonian
(\ref{eq:Dirac1})-(\ref{eq:Dirac2}) with only a gauge field $\vec{A}$:
\begin{eqnarray}
  \label{eq:dirac-gauge}
 - i v_0 \vec{\sigma}\cdot \left(\vec{\nabla} +  e \vec{A}
  \right)  \Psi = E \Psi .
\end{eqnarray}
It is known from vector analysis that any vector can be
decomposed into the sum of a gradient and a rotational. Using that
property, we can express the 2D vector $\vec{A}$ as
\begin{eqnarray}
  \label{eq:decomposition}
  \vec{A}=\vec{\nabla} \chi + (\hat{z} \times \vec{\nabla}) \phi .
\end{eqnarray}
Using this decomposition, we can rewrite $\vec{\sigma}\cdot \vec{A}$
in the form
\begin{eqnarray}
  \vec{\sigma}\cdot \vec{A} = \vec{\sigma}\cdot \vec{\nabla} \chi +
  \vec{\sigma}\cdot (\hat{z} \times \vec{\nabla}) \phi .
\end{eqnarray}
The mixed product $\vec{\sigma}\cdot (\hat{z} \times
\vec{\nabla} \phi)=(\vec{\sigma} \times \hat{z}) \cdot\vec{\nabla} \phi$,
and $i \vec{\sigma} \sigma_z= (\vec{\sigma} \times \hat{z})$ so the
Dirac equation (\ref{eq:dirac-gauge}) can be rewritten as:
\begin{eqnarray}
  \label{eq:dirac-transformed}
  -v_0 \vec{\sigma}\cdot \left(\vec{\nabla} + e
      \vec{\nabla} \chi + i e \sigma_z \vec{\nabla} \phi
  \right)  \Psi = E \Psi .
\end{eqnarray}
Then, a pseudogauge transformation to a new wave function $\tilde{\Psi}$
according to
\begin{eqnarray}
  \label{eq:cov}
  \Psi = e^{e (i \chi -\sigma_z \phi)} \tilde{\Psi},
\end{eqnarray}
turns the Dirac equation (\ref{eq:dirac-transformed}) into the free Dirac
equation without vector potential
\begin{eqnarray}
  \label{eq:simple}
  -i v \vec{\sigma}\cdot \vec{\nabla} \tilde{\Psi} = E \tilde{\Psi},
\end{eqnarray}
and thus  the transport properties of the initial model (\ref{eq:dirac-gauge}) turn out to be the same as in the absence
of the gauge potential.

However, there are some subtleties   in applying this argument to correlated
in space gauge potential. The difficulty
stems from the fact that the
transformation (\ref{eq:cov}) is not unitary. Indeed, if we denote the
original and the transformed wave functions by
\begin{equation}
 \Psi(x,y) =  \left(\begin{array}{c}
    u(x,y) \\ v(x,y)
  \end{array} \right), \ \ \
 \tilde{\Psi}(x,y) =  \left(\begin{array}{c}
    \tilde{u}(x,y) \\ \tilde{v}(x,y)
  \end{array} \right).
\end{equation}
Then, the normalization condition
\begin{equation}
  \label{eq:normalize}
  1 = \int dx dy [|u(x,y)|^2 +  |v(x,y)|^2],
\end{equation}
transforms under pseudogauge transformation to
\begin{equation}
  \label{eq:normalize-tilde}
  1 = \int dx dy [e^{-2 e \phi(x,y)} |\tilde{u}(x,y)|^2 + e^{ 2 e \phi(x,y)}
  |\tilde{v}(x,y)|^2].
\end{equation}
Therefore, the normalization condition (\ref{eq:normalize-tilde}) is
equivalent to the normalization condition (\ref{eq:normalize}) only
for extended states, and only when $\phi(x,y)$ is vanishing outside of
a finite area. Indeed, in that case the normalization
integral is dominated by the asymptotic behavior
of the extended wavefunctions outside the finite area and is thus
unchanged by the transformation.

Thus, if $\phi(x,y)$ is vanishing outside of a finite region, this
would also imply a vanishing gradient and thus vanishing correlations of the
vector potential outside of this region.
As a result, the correlations of the vector
potential become necessarily finite ranged, in contradiction with the
hypothesis of an infinite ranged power-law decay. This is in contrast
with the case of a $\delta$-correlated gauge potential which is
compatible with a potential existing only in a finite region of space.
Nevertheless, we have not found any corrections to transport to one-loop order.

\subsection{Random mass disorder}

Let us now consider the system with only random mass disorder. The corresponding
flow equations are
\begin{eqnarray}
\frac{\partial\tilde{\alpha}_z}{\partial \ln L } &=& - 2 \tilde{\alpha}_z^2
-2\tilde{\alpha}_z\tilde{\beta}_z,  \\
\frac{\partial\tilde{\beta}_z}{\partial \ln L } &=& (2-a)\tilde{\beta}_z -
2 \tilde{\beta}_z^2
-2\tilde{\alpha}_z\tilde{\beta}_z,  \\
\frac{\partial \ln \tilde{\varepsilon}}{\partial \ln L } &=& 1+
\tilde{\alpha}_{z}+\tilde{\beta}_{z}.
\end{eqnarray}
In the case of SR correlated disorder the running disorder strength approaches
the Gaussian fixed point ($\tilde{\alpha}_z^*=0$) so that the SR disorder is
marginally irrelevant. This results in the logarithmic corrections to the
scaling of the  density of states
\begin{eqnarray}
\rho_{\mathrm{SR}}(\varepsilon)&=&\frac{\alpha_z\, \varepsilon}{\pi v_0^2}
\ln \frac{\Delta}{\varepsilon}. \label{eq:DOS-RM-1}
\end{eqnarray}
In the most general case
the flow equations possess beside the unstable
Gaussian fixed point ($\tilde{\alpha}_z^*=\tilde{\beta}_z^*=0$)  a nontrivial
infrared stable fixed point ($\tilde{\alpha}_z^*=0$, $\tilde{\beta}_z^*=(2-a)/2$)
with eigenvalues $\lambda_{1,2}=-2+a$ negative for $a<2$.
The dynamic exponent describing the energy scaling is then given by
\begin{eqnarray}
z=1+\tilde{\alpha}_z^*+\tilde{\beta}_z^* = 1+(2-a)/2.
\end{eqnarray}
The density of states has then universal scaling behavior
\begin{eqnarray}
\rho_{\mathrm{LR}}(\varepsilon)\sim \varepsilon ^{(2-z)/z}.
\end{eqnarray}
The system of 2D Dirac fermions (or more precisely the pair of
Majorana fermions)
with random mass disorder is formally equivalent to two decoupled
classical 2D Ising models with random bond disorder at
criticality.\cite{dotsenko83}
It is known that uncorrelated random bond disorder is irrelevant in RG sense
resulting only in logarithmical corrections to the scaling of the pure Ising
model. However, the LR correlated disorder is a relevant perturbation
that changes
the critical behavior.\cite{rajabpour08} This latter result is in accordance with our
findings.

The conductance and the Fano factor at the Dirac cone given by
\begin{eqnarray}
G&=&\frac{e^2}{\pi h}\frac{W}{L}\left[1- f_{z}'(0)(2-a)/2\
 \right],  \label{eq-mass-1} \\
 F&=& \frac13  +
 \frac23 \frac{f'''_{z}(0)(2-a)/2
 }{1-f'_{z}(0)(2-a)/2}. \label{eq-mass-2}
\end{eqnarray}
The conductance and the Fano factor turn out to be also universal. They are
shown in Figs.~(\ref{fig:con}) and
(\ref{fig:fano}) as a functions of $a$. Since upon renormalization the disorder
couplings approach a fixed point of order $2-a$ the system does not develop the
mean free path scale. Thus, one can expect that the expressions for
conductance (\ref{eq-mass-1}) and the Fano factor (\ref{eq-mass-2}) hold
up to very large scale. Remarkably, that in contrast to
uncorrelated disorder which suppresses the Fano factor the correlated
disorder  can enhance it.
In the case of adatoms on the surface of topological insulator undergoing
the paramagnetic-ferromagnetic Ising-like phase transition with $\eta=a=1/4$
one would expect on the basis of our one-loop treatment that the
density  of surface states behaves as
$\rho(\varepsilon)\sim \varepsilon^{1/15}$.
Unfortunately,  for $a \ll 2$ the lowest order
correction in disorder becomes too large so that one cannot
rely anymore on the one-loop approximation.

\section{Random gauge potential: bosonisation}
 \label{sec:bosonization}
In this section we reanalyze the problem of 2D Dirac fermions in the presence
of LR correlated random gauge potential  with the bosonisation
technique. We will first give a detailed derivation of the bosonized
action in the case of a general interaction,  then discuss first
the SR correlated disorder case\cite{ludwig94,tsvelik94,nersesyan95}
before turning to the LR correlated case and comparing our results with
those of Sec.~\ref{sec:RG}.
We start from the replicated action~(\ref{eq:action}) with only terms
with  $\mu=x,y$. In the partition function path integral, we introduce in
the action the Matsubara time variable
$\tau=y/v_0$ and we make the change of (independent) Grassmann
variables according to:
\begin{eqnarray}
  \label{eq:cov-grassmann}
  \bar{\psi}_a= \tilde{\psi}^\dagger_a \frac{i\sigma_y}{\sqrt{v_0}}, \ \ \ \  \
  \psi_a =  \frac{\tilde{\psi}_a}{\sqrt{v_0}}.
\end{eqnarray}
The transformed action $S = S_0+V$,
which we split for convenience into a free and
interacting part, reads
\begin{eqnarray}
  \label{eq:new-action-0}
  S_0&=&\int dx d\tau \sum_a \psi_a^\dagger [-\partial_\tau + i v_0
  \sigma_z \partial_x -\nu_n \sigma_y ] \psi_a, \\
   V&=& \frac12 \pi v_0^2 \int dx\, dx'\, d\tau\, d\tau' g(x-x',\tau-\tau') \nonumber \\
 &&\!\!\!\!\!\!\!  \left[ \left(\sum_a ( \psi^\dagger_a
   \sigma_z \psi_a)(x,\tau)\right)\left( \sum_b ( \psi^\dagger_b
   \sigma_z \psi_b)(x',\tau')  \right) \right. \nonumber \\
   &&\!\!\!\!\!\!\!\!\! \left.-  \left(\sum_a ( \psi^\dagger_a
    \psi_a)(x,\tau)\right)\left( \sum_b ( \psi^\dagger_b
    \psi_b)(x',\tau')  \right) \right], \ \ \ \label{eq:new-action-int}
\end{eqnarray}
where we have dropped the tildes for clarity. The function $g(x,\tau)$
for SR correlated disorder is given by
$g(x,\tau)=\alpha_{\perp}\delta(x)\delta(v_0\tau)$ and for LR correlated disorder
by $g(x,\tau)=\beta_{\perp}\mathcal{A}_a (x^2+v_0^2 \tau^2)^{-a/2}$.
This action has the form of the action of a model of interacting
fermions with an interaction that is non-local in Matsubara
time.\cite{negele_book} The density of states can be calculated as
\begin{eqnarray}
  \label{eq:density-of-states}
  \rho(\varepsilon)=-\frac 1 \pi\  \mathrm{Im}\,  \mathrm{Tr} \mathcal{G}(i\nu_n \to
  \varepsilon+i0_+)
\end{eqnarray}
with the trace of the Matsubara Green function given by
\begin{eqnarray}
  \label{eq:green}
 \mathrm{Tr} \mathcal{G}(i\nu_n)&=&\mathrm{Tr}[(i\nu_n + \mu -H)^{-1}],
 \nonumber \\
           &=& \frac{\partial}{\partial (i\nu_n)} [\ln Z(i\nu_n)],
           \nonumber \\
           &=& \frac i {v_0} \langle \psi^\dagger \sigma_y \psi \rangle,
\end{eqnarray}
where $H$ is the corresponding Hamiltonian and $Z$ the partition function.
In Eq.~(\ref{eq:green}), the average is taken with respect to the
action~(\ref{eq:new-action-0})--~(\ref{eq:new-action-int}).
It is convenient for the bosonization procedure to introduce the components:
\begin{equation}
  \label{eq:components-dirac}
  \psi_a(r)=\left(
    \begin{array}{c}
      \psi_{R,a} \\ \psi_{L,a}
    \end{array}
\right), \ \ \  \
  \psi^\dagger_a(r)=\left(
    \begin{array}{c}
      \psi^\dagger_{R,a} \\ \psi^\dagger_{L,a}
    \end{array}
\right)
\end{equation}
and define:
\begin{eqnarray}
 J_L = \sum_a \psi^\dagger_{L,a} \psi_{L,a}, \  \  \  \
 J_R = \sum_a \psi^\dagger_{R,a} \psi_{R,a}
\end{eqnarray}
to rewrite the interacting part of the action in the form:
\begin{eqnarray}
  \label{eq:bosonized-interacting}
  V&=& -\pi v_0^2 \int dx\, dx'\, d\tau\, d\tau'\, g(x-x',\tau-\tau') \nonumber \\
  && \times [J_R(x,\tau) J_L(x',\tau') +  J_L(x,\tau) J_R(x',\tau')].\ \ \ \
\end{eqnarray}
We can now apply the bosonization technique to the action
(\ref{eq:new-action-0})-(\ref{eq:new-action-int}). First, the
Hamiltonian of the non-interacting part is rewritten in terms of the
components~(\ref{eq:components-dirac})  as:
\begin{eqnarray}
  H_0&=&\sum_a H_{0,a} \\
  H_{0,a} &=& \int dx  \left[-i v_0  (\psi^\dagger_{R,a} \partial_x
  \psi_{R,a} - \psi^\dagger_{L,a} \partial_x \psi_{L,a}) \right. \nonumber \\
  && \left.  + i\nu_n
  (\psi^\dagger_{R,a} \psi_{L,a} -\psi^\dagger_{L,a} \psi_{R,a})\right].
\end{eqnarray}
In bosonization, the fermion fields are expressed in terms of bosonic
fields\cite{schulz_houches_revue,giamarchi_book_1d}
$\theta_a$ and $\phi_a$ as:
\begin{eqnarray}
  \psi_{R,a} &=& \frac{1}{\sqrt{2\pi \Lambda}} e^{i(\theta_a - \phi_a)}
  \eta_{R,a} \\
  \psi_{L,a} &=& \frac{1}{\sqrt{2\pi \Lambda}} e^{i(\theta_a + \phi_a)}
  \eta_{L,a},
\end{eqnarray}
with $\Lambda$ a short distance cutoff,
$\partial_x \theta_a = \pi \Pi_a$ and  the fields $\phi_a$ and
$\Pi_a$ satisfy the canonical commutation relations
$[\phi_a(x),\Pi_b(x')]=i\delta_{ab} \delta(x-x')$. The operators
$\eta_{R/L,a}$
are Majorana fermion operators that ensure anticommutation of the
fermion fields. The disorder-free part of the Hamiltonian has the
bosonized form\cite{giamarchi_book_1d}
\begin{eqnarray}
  H_{0,a} &=& \int \frac{dx}{2\pi} \left[(\pi \Pi_a)^2 + (\partial_x
  \phi_a)^2\right] \nonumber \\
  && - \frac{\nu_n}{2\pi \Lambda} \int dx \cos 2\phi_a,
\end{eqnarray}
where we have chosen the same eigenvalue $-i$
for all the products $\eta_{R,a} \eta_{L,a}$.
The disorder-free  part of the action is then:
\begin{eqnarray}\label{eq:bosonized-no-disorder}
  S_0 = i \sum_a \int dx d\tau \Pi_a \partial_\tau \phi_a - \int d\tau
  H_0.
\end{eqnarray}
After integrating out the fields $\Pi_a$ in the
path integral with action~(\ref{eq:bosonized-no-disorder}),
the action of the sine-Gordon model is
obtained.\cite{rajaraman_instanton}
In the presence of disorder, we introduce the symmetric combinations of
the bosonic fields:
\begin{eqnarray}
  \phi_C = \frac 1 {\sqrt{n}} \left(\sum_a \phi_a \right), \  \  \
  \Pi_C =  \frac 1 {\sqrt{n}} \left(\sum_a \Pi_a \right),
\end{eqnarray}
and the new fields $\phi_\lambda,\Pi_\lambda$  with $1\le
\lambda \le n-1$ such that:
\begin{eqnarray} \label{eq:transf-replica}
  \phi_a = \frac{\phi_c}{\sqrt{n}} +  \sum_{\lambda=1}^{n-1}
  e_a^\lambda \phi_\lambda \\
\Pi_a = \frac{\Pi_c}{\sqrt{n}} +  \sum_{\lambda=1}^{n-1}
  e_a^\lambda \Pi_\lambda
\end{eqnarray}
with:
\begin{eqnarray}\label{eq:ortho-replica}
  \frac 1 n  + \sum_{\lambda=1}^{n-1} e_a^\lambda e_b^\lambda =
  \delta_{ab}, \nonumber \\
  \sum_{a=1}^n e_a^\lambda =  0,  \nonumber \\
  \sum_{a=1}^n e_a^\lambda e_a^\mu = \delta_{\lambda \mu}.
\end{eqnarray}
The conditions (\ref{eq:ortho-replica}) ensure that the new fields
defined in Eq.~(\ref{eq:transf-replica}) satisfy the canonical
commutation relations. We can then express the disorder
contribution~(\ref{eq:bosonized-interacting})
to the action entierely in terms of $\Pi_C$ and $\phi_C$ thanks to the
relations:
\begin{eqnarray}
  J_{R/L}=-\frac{\sqrt{n}}{2\pi} \partial_x \phi_C \pm \frac{\sqrt{n}}{2} \Pi_C.
\end{eqnarray}
We will now discuss separately the two cases of SR and LR correlated
disorder.

\textit{SR correlated disorder.}
In the case of $g(x,\tau)=\alpha_{\perp}\delta(x)\delta(v_0\tau)$, the
disordered part of the action (\ref{eq:bosonized-interacting}) can be
rewritten as
\begin{eqnarray}
  V=- 2\pi n \alpha_{\perp} v_0 \int dx d\tau \left[ \frac{(\partial_x
      \phi_C)^2}{4\pi^2} - \frac{\Pi^2_C}{4}\right].
\end{eqnarray}
The fields $\Pi_C$ and $\Pi_\lambda$ are then  integrated out,
leaving an action expressed purely in terms of  $\phi_C,\phi_\lambda$.
The quadratic part of the action
of the fields $\phi_\lambda$ is unchanged compared with the case without
disorder, but the action of the field $\phi_C$ becomes:
\begin{eqnarray}
  \int \frac{dx d\tau}{2\pi} \left[\frac{(\partial_\tau \phi_C)^2}{v_0(1-n\alpha_{\perp} )}
  + v_0(1+n\alpha_{\perp} ) (\partial_x \phi_C)^2\right].
\end{eqnarray}
The common scaling dimension of the fields $\cos 2\phi_a$ then becomes:
\begin{eqnarray}
  \mathrm{dim.}(\cos 2\phi_a) = 1 +\frac{K_C-1} n,
\end{eqnarray}
where
\begin{eqnarray}
  K_C=\sqrt{\frac{1 - n\alpha_{\perp}}{1 + n\alpha_{\perp}}},
\end{eqnarray}
and for $n \to 0$,
\begin{eqnarray}\label{eq:scal-dim-cos}
   \mathrm{dim.}(\cos 2\phi_a) \to 1 - \alpha_{\perp},
\end{eqnarray}
leading to the following renormalization group equation
\begin{eqnarray}
  \frac{d \nu_n}{d\ell} = \left(1 +  \alpha_{\perp} \right)
  \nu_n
\end{eqnarray}
for $\nu_n$. A strong coupling scale is reached for $e^{\ell^*}\sim
|\nu_n\Lambda/v_0|^{-1/(1+\alpha_\perp)}$. Using the
Eq.~(\ref{eq:green}) and the scaling dimension~(\ref{eq:scal-dim-cos})
the density of states is obtained as:\cite{ludwig94,tsvelik94,nersesyan95}
\begin{eqnarray}\label{eq:boso-dos-SR}
  \rho_{\mathrm{SR}}(\varepsilon) =\frac{1}{2\pi \Lambda v_0}
  \left(\frac{\varepsilon\Lambda}{v_0}\right)^{\frac{1 -
   \alpha_{\perp}}{1 +  \alpha_{\perp}}},
\end{eqnarray}
\emph{i.e} a power-law enhancement with non-universal exponent is
obtained. By comparing Eq.~(\ref{eq:boso-dos-SR}) with the RG
caculation result of Eq.~(\ref{eq:DOS-RG-SR})  we note that the two
results are in perfect agreement provided the short distance cutoff
is taken as $\Lambda=v_0/\Delta$.

\textit{LR correlated disorder.}
In this case, introducing the Fourier transform
 $\hat{g}(q,\omega)=\frac {\beta_\perp} {v_0} (q^2 +
 \omega^2/v_0^2)^{(a-2)/2}$ of $g(x,\tau)$,
we rewrite the action as:
\begin{eqnarray}\label{eq:LR-action}
  S&=& \int \frac{dq d\omega}{2\pi^2} \frac{|\phi_C(q,\omega)|^2}{2\pi v_0}
  \left[ \frac{\omega^2}{1- n v_0 \hat{g}(q,\omega)} + (v_0
    q)^2 \left(1+\right.\right. \nonumber\\
&& \left. \left.     n v_0 \hat{g}(q,\omega)\right)\right]
 + \sum_{\lambda} \int \frac{dx d\tau}{2\pi} \left[
    \frac{(\partial_\tau \phi_\lambda)^2}{ v_0} + v_0 (\partial_x
    \phi_\lambda)^2 \right] \nonumber \\
&& -\frac {\nu_n}{2\pi \Lambda} \sum_n \int dx d\tau \cos 2 \phi_n.
\end{eqnarray}
In general, a model with an action such as (\ref{eq:LR-action})
is not integrable. To estimate the free energy associated with
(\ref{eq:LR-action}) we use the Gaussian Variational
Method\cite{mezard_variational_replica} with (replica symmetric)
variational  action:
\begin{eqnarray}
  S_{\mathrm{var}}&=& \int \frac{dq d\omega}{2\pi^2}
  \frac{|\phi_C(q,\omega)|^2}{2\pi v_0}
  \left[ \frac{\omega^2}{1- n v_0 \hat{g}(q,\omega)} + \right. \nonumber\\
&&  (v_0
    q)^2 \left(1+
 \left.     n v_0 \hat{g}(q,\omega)\right)\right]\nonumber\\
&& + \sum_{\lambda} \int \frac{dx d\tau}{2\pi} \left[
    \frac{(\partial_\tau \phi_\lambda)^2}{ v_0} + v_0 (\partial_x
    \phi_\lambda)^2 \right] \nonumber \\
&& + \frac{\omega_0^2}{2\pi v_0} \sum_a \phi_a^2,
\end{eqnarray}
and minimize the variational free energy:
\begin{eqnarray}
  F_{\mathrm{var}}&=&F_0+\langle S-S_{\mathrm{var}}\rangle_{S_{\mathrm{var}}}, \\
  F_0&=&-\ln \left[\int \prod_a {\mathcal D} \phi_a e^{-S_{\mathrm{var}}} \right].
\end{eqnarray}
After some calculation, we find that:
\begin{eqnarray}
  \label{eq:variational-condition}
&&  \omega_0^2 = \frac{|\nu_n| v_0}\Lambda e^{-2 \langle \phi_a^2 \rangle },
 \nonumber  \\
&& \lim_{n\to 0} \langle \phi_a^2 \rangle = \frac 1 2 \ln \left(\frac {v_0}
   {\Lambda \omega_0}\right)\nonumber \\
&& -\pi v_0 \int \frac{d\omega d(v_0q)}{4\pi^2}
 \frac{\omega^2 + (v_0q)^2} {(\omega^2 + (v_0 q)^2+\omega_0^2)^2}
 \hat{g}(q,\omega).\ \ \ \
\end{eqnarray}
Solving the self-consistent equation (\ref{eq:variational-condition}),
we obtain:
\begin{eqnarray}
  \label{eq:boso-LR-mass}
&&    \frac{\beta_\perp}{\zeta(a)} \left(\frac {\omega_0} {v_0}
  \right)^{a-2}  = W \left[ \frac{\beta_\perp}{\zeta(a)} \left(\frac {|\nu_n|} {v_0}
  \right)^{a-2} \right], \nonumber
\end{eqnarray}
where $W(x)$ is the already appeared in Sec.~\ref{sec:SCBA}
Lambert function\cite{olver2010} and we have introduced the function
\begin{equation}
\zeta(a)= \frac{8}{\pi a (2-a)}  \sin \left(\frac{\pi  a}{2}\right).
\end{equation}
We find a density of states $\rho(\nu_n)=\frac{\omega_0^2(\nu_n)}{2\pi
  v_0^2 |\nu_n|}$ that behaves for low energy as:
\begin{eqnarray}
\rho_{\mathrm{LR}}(\varepsilon)=\frac{1}{2\pi \varepsilon}
\left(\frac{\zeta(a)}{{\beta}_{\bot}} \ln \frac{\Delta}{\varepsilon}
\right)^{-2/(2-a)}, \label{eq:DOS-gauge-2}
\end{eqnarray}
hence, the density of state has a divergence for $\varepsilon \to 0$ which
is, however, integrable. Note that the result~(\ref{eq:DOS-gauge-2})
is independent from the cutoff $\Lambda$.
The density of states~(\ref{eq:DOS-gauge-2})
is expected to be valid down to zero energy and agrees with the prediction
of RG~(\ref{eq:DOS-gauge-1}) which is supposed to be valid at energies larger
than the exponentially small in the limit $a\to2$ energy scale. This proves
that there exists only a single regime with the asymptotic
behavior~(\ref{eq:DOS-gauge-2}).

Note, that the result~(\ref{eq:DOS-gauge-2}) differs from
the density of states
$\rho(\varepsilon)\sim 1/(\varepsilon \ln |\varepsilon|^{(6-a)/(2-a)})$
obtained in Ref.~\onlinecite{khveshchenko08} using a supersymmetric
approach and a variational approximation.
We found that Eq.~(20) of Ref.~\onlinecite{khveshchenko08}
is not a correct solution of Eq.~(19) of that paper. Upon finding the correct
solution the subsequent calculations reproduce our result~(\ref{eq:DOS-gauge-2}).

\section{Conclusions}
\label{sec:summary}

We have studied 2D Dirac fermions in the presence of LR correlated disorder
with correlations decaying as a power law. In particular we
have considered three types of disorder: random scalar potential, random gauge
potential and random mass. Using the SCBA, weak
disorder RG and bosonisation technique we have computed the density of states
modified by disorder in vicinity of the Dirac point of free fermions.
Using a diagrammatic technique with matrix Green functions we have derived
the full counting statistics of fermionic transport at low energy.
Remarkably, in contrast with SR correlated disorder the LR correlated disorder
provides deviation from the pseudodiffusive transport already to lowest order
in disorder.

In the case of LR
correlated random potential the picture resembles that for SR correlated random
potential.  Using the SCBA and RG
give qualitatively consistent picture: disorder generates an algebraically small
energy scale below which the density of states saturates to a constant value
while  above this scale it is given by a corrected bare density of states.
The correction to the conductance due to LR correlated disorder at the Dirac cone
is positive and increases with $a$ while the correction to the Fano factor
is small and negative.

For LR correlated random gauge potential we have found that the density of states
diverges at zero energy in an integrable way. This small energy behavior
derived using bosonisation is completely consistent with the prediction of RG
which is valid for larger energies. In particular, the density of states
is accessible in graphene using STM measurements that would allow one
to measure the real exponent $a$ describing the correlation of the random gauge
potential induced by ripples.
We have found that the LR correlated random gauge potential does
not contribute to the transport properties to one-loop order.

In the case of LR correlated random mass disorder we have found a non-trivial
infrared stable fixed point  which controls the large scale properties of the
disordered Dirac fermions. This results in a universal power law behavior of
the density of states and universal transport properties. Since the disorder couplings
flow to the fixed point the system does not exhibit the mean free path scale.
Thus, the conductivity and the Fano factor at the Dirac point are expected to
have universal forms up to very large scales. Remarkably, that in contrast to
uncorrelated disorder which suppresses the Fano factor the correlated
random mass disorder enhances it.

\acknowledgments

We acknowledge the support from  ANR through the grant 2010-Blanc IsoTop.

\appendix
\section{One-loop diagrams contributing to the free energy}
\label{sec:integrals}

In this appendix we compute the diagrams shown in Fig.~\ref{fig:free-energy}(a)
with the dashed line corresponding to 3 different disorder correlators.
To that end we substitute the bare Green function~(\ref{eq:G-Naz-1})
in Eq.~(\ref{eq:diag-naz}) and evaluate the trace explicitly.
Since the diagrams contain $\phi$ independent divergent terms we will compute
the derivatives of the diagrams with respect to $\phi$.
The diagrams with LR correlated scalar and random mass disorder then yield
\begin{eqnarray}
f_{0,z}(\phi)&=&
\int\limits_{0}^{\infty}dy \int\limits_0^{1}
 dx_1\int\limits_0^{1} dx_2
 \frac{2 \pi ^2 \mathcal{A}_a  y \sinh (y \phi )}{\left(y^2+(x_1-x_2)^2\right)^{a/2}}
  \nonumber \\
&&\times  \left(  \frac{1}{\cosh (\pi y) - \cos (\pi(x_1-x_2))}  \right. \nonumber \\
&& \left. \mp \frac{1}{\cosh(\pi  y)-\cos (\pi  (x_1+x_2))}   \right),
\label{eq:f0-old}
\end{eqnarray}
where the upper sign corresponds to $f_0$ and the lower sign to $f_z$.
The diagram with LR correlated random gauge disorder gives an expression which
does not depend on $\phi$ and thus it does not contribute to transport.
We now change variables from $x_1$ and $x_2$ such that $\cos(\pi(x_1+x_2))=b$
and $x_2-x_1=c$ that formally can be written as
\begin{eqnarray}
&&\int\limits_0^1d x_1 \int \limits_0^1d x_2 f(\cos[\pi(x_1+x_2)],|x_2-x_1|) \nonumber \\
&& = \int\limits_0^1d c \int \limits_{-1}^{\cos \pi c} \frac{2db}{\sqrt{1-b^2}}
f(b,c). \label{eq:change-var}
\end{eqnarray}
Applying transformation~(\ref{eq:change-var}) to Eq.~(\ref{eq:f0-old})
and evaluating the integration over $b$ we obtain Eq.~(\ref{eq:fz}).


\end{document}